%% file: 2018Team1FMTC-BR-Report.tex
\documentclass[preprint,authoryear,11pt]{report}

\usepackage{tcolorbox}
\tcbuselibrary{theorems}

\newtcbtheorem{mytheo}{}%
{colback=green!5,colframe=green!35!black,fonttitle=\bfseries}{th}

\usepackage{amsmath}
\usepackage{amsthm}
\usepackage{amssymb}
\usepackage{graphicx}
\usepackage[margin=1.5in]{geometry}
\usepackage{pdflscape}
\usepackage{palatino}
\usepackage{hyperref}
\usepackage{natbib}
\usepackage{titlepic}
\usepackage{graphicx}
\usepackage{setspace}
\usepackage{float}
\usepackage{verbatim}
\usepackage{tikz}
\usetikzlibrary{snakes}
\usepackage[framed]{mcode}
\usepackage{amsmath,varwidth,array,ragged2e}
\newcolumntype{L}{>{\varwidth[c]{\linewidth}}l<{\endvarwidth}}
\newcolumntype{M}{>{$}l<{$}}
\usepackage{units}

\graphicspath{ {images/} }
\usepackage{xcolor}
\usepackage{lipsum}
\definecolor{titlepagecolor}{cmyk}{1,.60,0,.40}
\definecolor{namecolor}{cmyk}{1,.50,0,.10} 

\theoremstyle{definition}

\newcommand{\bigO}{\mathcal{O}}

\usepackage[latin1]{inputenc}
\usepackage{amsmath}
\usepackage{amsfonts}
\usepackage{amssymb}
\usepackage{graphicx}
\usepackage{color}
\usepackage{tikz}

\newcommand{\Ito}{It\^{o} }

\newcommand{\PP}{{\mathbb P}}
\newcommand{\QQ}{{\mathbb Q}}
\newcommand{\EE}{{\mathbb E}}
\newcommand{\RR}{{\mathbb R}}
\newcommand{\LL}{{\mathcal L}}

\newcommand{\As}{{\mathcal A}}
\renewcommand{\vec}[1]{\mathbf{#1}}
\newcommand{\Ls}{{\mathcal{L}}}

\newcommand{\vx}{{\boldsymbol x}}
\newcommand{\vy}{{\boldsymbol y}}
\newcommand{\vb}{{\mathbf b}}
\newcommand{\vw}{{\mathbf w}}
\newcommand{\vu}{{\mathbf u}}
\newcommand{\vtheta}{{\boldsymbol \theta}}
\newcommand{\bx}{{\mathbf x}}
\newcommand{\by}{{\mathbf y}}
\newcommand{\vX}{{\boldsymbol X}}

\usepackage[labelfont=bf]{caption} 
\usepackage{afterpage}
\usepackage{footnote}
\usepackage{url}
\newcommand\redunderbrace[2]{{\color{red} \underbrace{\color{black} #1}_{\tiny \text{#2}} }}

\usepackage{graphicx}
\usepackage{wrapfig}
\usepackage{lscape}
\usepackage{rotating}
\usepackage{epstopdf}

\usepackage{subfigure}
\usepackage{hyperref}
\hypersetup{
	colorlinks=true,
	linkcolor=blue,
	filecolor=magenta,      
	urlcolor=cyan,
	citecolor = blue
}

\usepackage{titlesec}

\titleformat*{\section}{\LARGE\bfseries}
\titleformat*{\subsection}{\Large\bfseries}
\titleformat*{\subsubsection}{\large\bfseries}
\titleformat*{\paragraph}{\large\bfseries}
\titleformat*{\subparagraph}{\large\bfseries}

\usetikzlibrary{positioning}

\usepackage{chngcntr}

\usepackage[font={footnotesize,it}]{caption}
\usetikzlibrary{arrows.meta}
\usetikzlibrary{positioning}
\usetikzlibrary{calc}

\newcommand*{\titleGM}{\begingroup 
\hbox{ 
\hspace*{0.2\textwidth} 
\rule{1pt}{\textheight} 
\hspace*{0.05\textwidth} 
\parbox[b]{0.75\textwidth}{ 

\begin{flushleft}
{\noindent\Huge\bfseries Solving Nonlinear and High-Dimensional Partial Differential Equations via Deep Learning}\\[2\baselineskip] 
\end{flushleft}
{\large \textit{TEAM One }}\\[ \baselineskip] 
{\large{\textsc{Ali Al-Aradi}},  \hspace{0.65cm} \small{University of Toronto} \\ \large{\textsc{Adolfo Correia}}, \hspace{0.03cm} \small{Instituto de Matem\'{a}tica Pura e Aplicada}\\ \large{\textsc{Danilo Naiff}},  \hspace{0.62cm} \small{Universidade Federal do Rio de Janeiro}} \\ \large{\textsc{Gabriel Jardim}}, \hspace{0.29cm} \small{Funda\c{c}\~ao Getulio Vargas} \\ 
{\large{\textsc{\\\textit{Supervisor}}}: \\ \large{\textsc{Yuri Saporito}}, \hspace{0.52cm} \small{Funda\c{c}\~ao Getulio Vargas} \\ } 

\vspace{80mm} 
{\noindent \small{EMAp, Funda\c{c}\~ao Getulio Vargas, Rio de Janeiro, Brazil}}\\[\baselineskip] 
}}
\endgroup}

\begin{document}
	\restoregeometry 
 \thispagestyle{empty} 

\setcounter{tocdepth}{2}

\titleGM 

	\begingroup
		\makeatletter
		\def\@makeschapterhead#1{%
		  {\parindent \z@ \raggedright
		    \normalfont
		    \interlinepenalty\@M
		    \Huge \bfseries  #1\par\nobreak
		    \vskip 40\p@
		  }}
		\makeatother

	\endgroup

 	\tableofcontents
\newpage 
	 	
\input{Introduction.tex}
\newpage
\input{PartialDifferentialEquations.tex}

\newpage
\input{NumericalMethods.tex}
\newpage
\input{MachineLearning-NeuralNetworks.tex}
\newpage
\input{DGM.tex}
\newpage
\input{ourWork.tex}
\newpage

\bibliographystyle{elsarticle-harv}
\bibliography{references}
\newpage

\end{document}

%% file: Introduction.tex
\chapter{Introduction}

In this work we present a methodology for numerically solving a wide class of partial differential equations (PDEs) and PDE systems using deep neural networks. The PDEs we consider are related to various applications in quantitative finance including option pricing, optimal investment and the study of mean field games and systemic risk. The numerical method is based on the \textbf{Deep Galerkin Method} (DGM) described in \cite{sirignano2017dgm} with modifications made depending on the application of interest. \\

The main idea behind DGM is to represent the unknown function of interest using a deep neural network. Noting that the function must satisfy a known PDE, the network is trained by minimizing losses related to the differential operator, the initial/terminal conditions and the boundary conditions given in the initial value and/or boundary problem. The training data for the neural network consists of different possible inputs to the function and is obtained by sampling randomly from the region on which the PDE is defined. One of the key features of this approach is the fact that, unlike other commonly used numerical approaches such as finite difference methods, it is \textit{mesh-free}. As such, it does not suffer (as much as other numerical methods) from the curse of dimensionality associated with high-dimensional PDEs and PDE systems. \\ 

The \textbf{main goals} of this paper are to:

\begin{enumerate}
	\item Present a brief overview of PDEs and how they arise in quantitative finance along with numerical methods for solving them.
	\item Present a brief overview of deep learning; in particular, the notion of neural networks, along with an exposition of how they are trained and used.
	\item Discuss the theoretical foundations of DGM, with a focus on the justification of why this method is expected to perform well.  
	\item Elucidate the features, capabilities and limitations of DGM by analyzing aspects of its implementation for a number of different PDEs and PDE systems. \\
\end{enumerate}

\definecolor{cadmiumgreen}{rgb}{0.0, 0.42, 0.24}

\begin{figure}[h!]
	\centering 
\begin{tikzpicture}   

	\draw[step=0.5cm,black!30,very thin] (-2,-2) grid (2,2);

    \draw[red,thick] (-2.0,2.0) -- (-2.0,-2.0);
    \draw[blue,thick] (-2.0,-2.0) -- (2.0,-2.0);

    \draw[cadmiumgreen,thick] (0,0) -- (0,-0.5);
    \draw[cadmiumgreen,thick] (0,0) -- (-0.5,0);
   	\draw[cadmiumgreen,thick] (0,0) -- (0.5,0);


    \foreach \x in {-2.0,-1.5,...,2.0}{
    \foreach \y in {-2.0,-1.5,...,2.0}{
    \fill[black!80] (\x,\y) circle[radius=1.pt];
    }}
    
    \fill[cadmiumgreen] (0,0) circle[radius=1.3pt];
    
    \node at (-2.5,0) {\small $x$};    
    \node at (0,-2.5) {\small $t$};
    
    \node at (0,0.25) {\color{cadmiumgreen} \scriptsize $(t_i,x_j)$};
	
    \node[label={[align=center,label distance=0cm,text depth=-0.8cm]: \color{blue} \scriptsize initial }] at (1.875,-2.1) {}; 

    \node[label={[align=center,label distance=0cm,text depth=-0.8cm]: \color{blue}\color{blue} \scriptsize condition}] at (1.875,-2.37) {};
    
    \node[label={[align=center,label distance=0cm,text depth=0,rotate=90]: \color{red} \scriptsize boundary}] at (-2.37,1.875) {};

    \node[label={[align=center,label distance=0cm,text depth=0,rotate=90]: \color{red} \scriptsize condition}] at (-2.1,1.875) {};
    
    \node at (0,2.5) {\footnotesize mesh grid points};    
	
\end{tikzpicture} \qquad	\begin{tikzpicture}
	\node[inner sep=0pt] (russell) at (0,0)
	    {\includegraphics[width=0.415\textwidth]{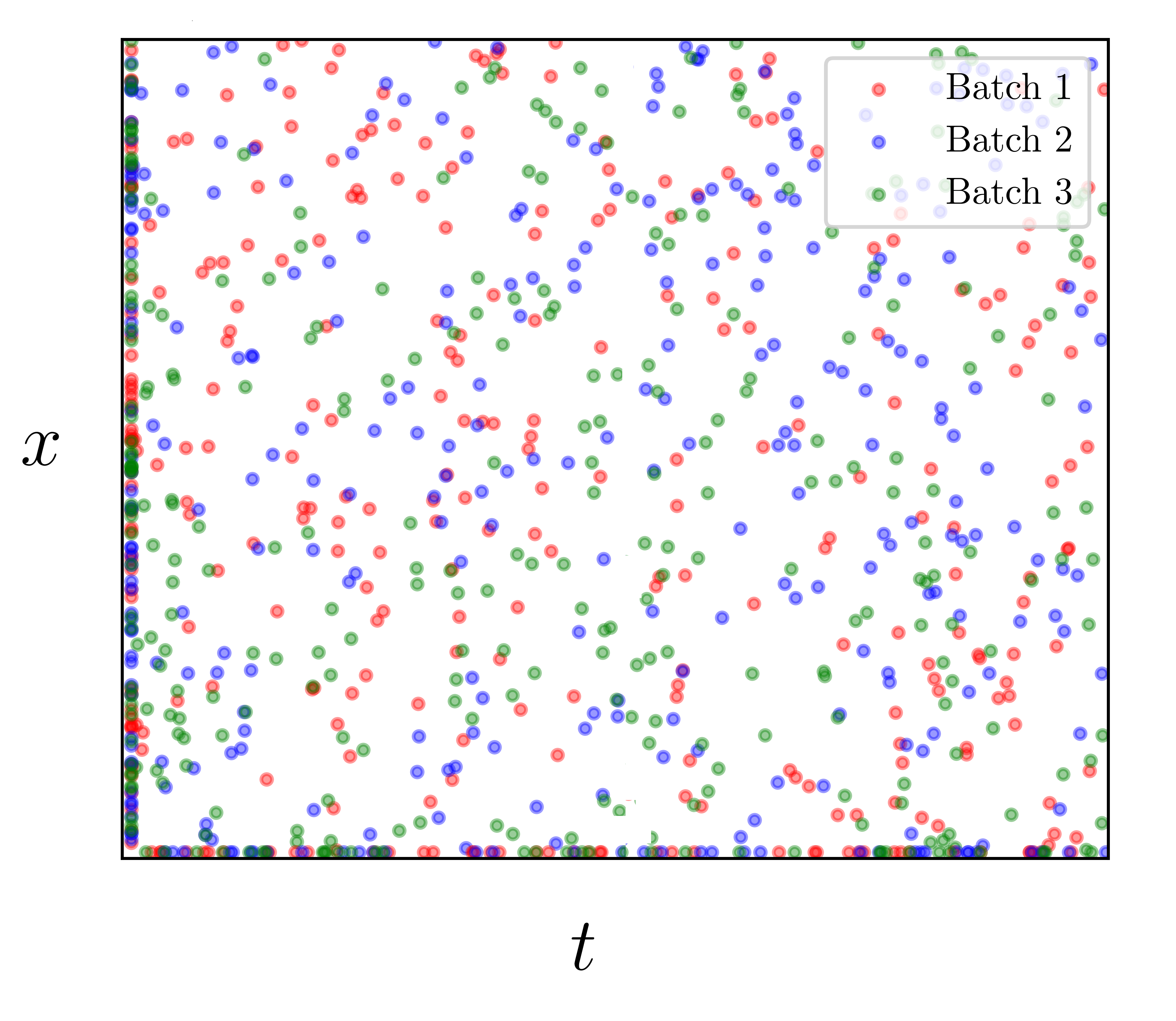}};
	\end{tikzpicture}
	\caption{Grid-based finite differences method (left) vs. Deep Galerkin Method (right)}\label{fig:DGMvsFD}
\end{figure}
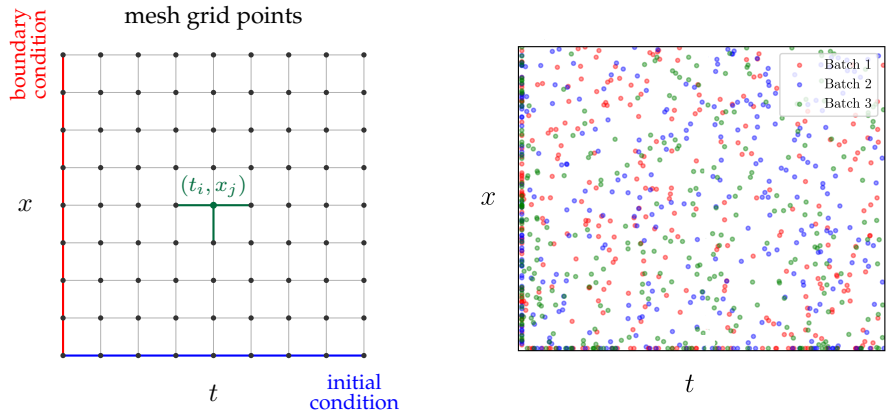

\vspace{0.7cm}

We present the results in a manner that highlights our own learning process, where we show our failures and the steps we took to remedy any issues we faced. The \textbf{main messages} can be distilled into \textbf{three main points}:
\begin{enumerate}
	\item \textbf{Sampling method matters}: DGM is based on random sampling; where and how the sampled random points used for training are chosen are the single most important factor in determining the accuracy of the method. 
	\item \textbf{Prior knowledge matters}: similar to other numerical methods, having information about the solution that can guide the implementation dramatically improves the results.
	\item \textbf{Training time matters}: neural networks sometimes need more time than we afford them and better results can be obtained simply by letting the algorithm run longer. \\ 
\end{enumerate}

%% file: PartialDifferentialEquations.tex
\chapter{An Introduction to Partial Differential Equations} \label{chap:PDE}
\setcounter{section}{2}

\subsection{Overview}

\textbf{Partial differential equations} (PDE) are ubiquitous in many areas of science, engineering, economics and finance. They are often used to describe natural phenomena and model multidimensional dynamical systems. In the context of finance, finding solutions to PDEs is crucial for problems of \textbf{derivative pricing}, \textbf{optimal investment}, \textbf{optimal execution}, \textbf{mean field games} and many more. In this section, we discuss some introductory aspects of partial differential equations and motivate their importance in quantitative finance with a number of examples.  \\
	
In short, PDEs describe a relation between a multivariable function and its partial derivatives. There is a great deal of variety in the types of PDEs that one can encounter both in terms of form and complexity. They can vary in \textbf{order}; they may be \textbf{linear} or \textbf{nonlinear}; they can involve various types of \textbf{initial/terminal conditions} and \textbf{boundary conditions}. In some cases, we can encounter \textbf{systems of coupled PDEs} where multiple functions are connected to one another through their partial derivatives. In other cases, we find \textbf{free boundary problems} or \textbf{variational inequalities} where both the function and its domain are unknown and both must be solved for simultaneously. \\

To express some of the ideas in the last paragraph mathematically, let us provide some definitions. A \textbf{$k$-th order partial differential equation} is an expression of the form:
\begin{equation*}
F\left( D^k u(x), D^{k-1}u(x), ... , D u(x), u(x), x \right) = 0 \, \qquad x \in \Omega \subset \RR^n
\end{equation*}
where $D^k$ is the collection of all partial derivatives of order $k$ and $u: \Omega \rightarrow \RR$ is the unknown function we wish to solve for. \\ 

PDEs can take one of the following forms:

\begin{enumerate}
\item \textbf{Linear PDE}: derivative coefficients and source term do not depend on any derivatives: 
\[ \sum_{|\alpha| \leq k} ~ \redunderbrace{a_\alpha(x) \cdot D^\alpha u}{\parbox{1.5cm}{\centering linear in \\ derivatives}} ~=~ \redunderbrace{f(x)}{\parbox{0.8cm}{\centering source \\ term}} \]
\item \textbf{Semi-linear PDE}: coefficients of highest order derivatives do not depend on lower order derivatives:
\[ \sum_{|\alpha| = k} ~ \redunderbrace{ \vphantom{a_0\left(D^{k-1}u ,..., Du, u, x \right)} a_\alpha(x) \cdot D^\alpha u}{\parbox{1.75cm}{\centering linear in \\ highest order \\ derivatives}} + \redunderbrace{a_0\left(D^{k-1}u ,..., Du, u, x \right)}{\parbox{3cm}{\centering source \\ term}} ~=~  0 \]
\item \textbf{Quasi-linear PDE}: linear in highest order derivative with coefficients that depend on lower order derivatives: 
\[ \sum_{|\alpha| = k} ~ \redunderbrace{a_\alpha \left(D^{k-1}u ,..., Du, u, x \right)}{\parbox{3.5cm}{\centering coefficient term of \\ highest order derivative}} \cdot D^\alpha u + \redunderbrace{a_0\left(D^{k-1}u ,..., Du, u, x \right)}{\parbox{3.85cm}{\centering source term does not depend \\ on highest order derivative}} = 0 \]
\item \textbf{Fully nonlinear PDE}: depends nonlinearly on the highest order derivatives. \\ 
\end{enumerate} 

A \textbf{system of partial differential equations} is a collection of several PDEs involving multiple unknown functions:
\begin{equation*}
\mathbf{F} \left( D^k \boldsymbol{u}(x), D^{k-1}\boldsymbol{u}(x), ... , D \boldsymbol{u}(x), \boldsymbol{u}(x), x \right) = 0 \qquad x \in \Omega \subset \RR^n
\end{equation*}
where $\boldsymbol{u}: \Omega \rightarrow \RR^m $. \\

\vspace{0.7cm}

Generally speaking, the PDE forms above are listed in order of increasing difficulty. Furthermore:
\begin{itemize}
	\item Higher-order PDEs are more difficult to solve than lower-order PDEs;
	\item Systems of PDEs are more difficult to solve than single PDEs;
	\item PDEs increase in difficulty with more state variables.
\end{itemize}
\vspace{0.5cm}

In certain cases, we require the unknown function $u$ to be equal to some known function on the boundary of its domain $\partial \Omega$. Such a condition is known as a \textbf{boundary condition} (or an \textbf{initial/terminal condition} when dealing with a time dimension). This will be true of the form of the PDEs that we will investigate in \hyperref[chap:DGM]{Chapter \ref{chap:DGM}}. \\


%


Next, we present a number of examples to demonstrate the prevalence of PDEs in financial applications. Further discussion of the basics of PDEs (and more advanced topics) such as well-posedness, existence and uniqueness of solutions, classical and weak solutions and regularity can be found in \cite{evans10}. \\

\setcounter{section}{2}

\subsection{The Black-Scholes Partial Differential Equation} 

One of the most well-known results in quantitative finance is the \textbf{Black-Scholes Equation} and the associated \textbf{Black-Scholes PDE} discussed in the seminal work of \cite{black1973pricing}. Though they are used to solve for the price of various financial derivatives, for illustrative purposes we begin with a simple variant of this equation relevant for pricing a European-style contingent claim. \\

\subsubsection{European-Style Derivatives}  
European-style contingent claims are financial instruments written on a source of uncertainty with a payoff that depends on the level of the underlying at a predetermined maturity date. We assume a simple market model known as the \textbf{Black-Scholes model} wherein a risky asset follows a geometric Brownian motion (GBM) with constant drift and volatility parameters and where the short rate of interest is constant. That is, the  dynamics of the price processes for a risky asset $X = (X_t)_{t \geq 0}$ and a riskless bank account $B = (B_t)_{t \geq 0}$ under the ``real-world'' probability measure $\PP$ are given by:
\begin{align*}
\frac{dX_t}{X_t} ~&=~ \mu ~dt + \sigma ~dW_t \\
\frac{dB_t}{B_t} ~&=~ r ~ dt
\end{align*} 
where $W = (W_t)_{t \geq 0}$ is a $\PP$-Brownian motion. \\

We are interested in pricing a claim written on the asset $X$ with payoff function $G(x)$ and with an expiration date $T$. Then, assuming that the claim's price function $g(t,x)$ - which determines the value of the claim at time $t$ when the underlying asset is at the level $X_t = x$ - is sufficiently smooth, it can be shown by dynamic hedging and no-arbitrage arguments that $g$ must satisfy the \textbf{Black-Scholes PDE}:

\begin{equation*}
\begin{cases}
\partial_t g(t,x) + r x \cdot \partial_x g(t,x)+ \frac{1}{2} \sigma^2 x^2\cdot \partial_{xx} g(t,x)  ~=~  r \cdot g(t,x) 
\\
g(T,x) ~=~ G(x)
\end{cases}
\end{equation*} \\

This simple model and the corresponding PDE can extend in several ways, e.g.
\begin{itemize}
	\item incorporating additional sources of uncertainty;
	\item including \textit{non-traded} processes as underlying sources of uncertainty;
	\item allowing for richer asset price dynamics, e.g. jumps, stochastic volatility;
	\item pricing more complex payoffs functions, e.g. path-dependent payoffs. \\
\end{itemize}

\subsubsection{American-Style Derivatives}
\label{sec:AmOpt}

In contrast to European-style contingent claims, American-style derivatives allow the option holder to exercise the derivative \textit{prior} to the maturity date and receive the payoff immediately based on the prevailing value of the underlying. This can be described as an \textbf{optimal stopping problem} (more on this topic in \hyperref[sec:optimalControl]{Section \ref{sec:optimalControl}}). \\

To describe the problem of pricing an American option, let $\mathcal{T}[t,T]$ be the set of admissible stopping times in $[t,T]$ at which the option holder can exercise, and let $\QQ$ be the risk-neutral measure. Then the price of an American-style contingent claim is given by: 

\[ g(t,x) = \underset{\tau \in \mathcal{T}[t,T]}{\sup} ~ \EE^\QQ \left[e^{-r(\tau - t)} G(X_\tau) ~\middle|~ X_t = x \right] \] 

Using dynamic programming arguments it can be shown that optimal stopping problems admit a \textbf{dynamic programming equation}. In this case, the solution of this equation yields the price of the American option. Assuming the same market model as the previous section, it can be shown that the price function for the American-style option $g(t,x)$ with payoff function $G(x)$ - assuming sufficient smoothness - satisfies the \textbf{variational inequality}:  

\begin{equation*}
\max \left\{ \left(\partial_t + \LL - r \right) g , ~ G - g \right\} = 0, \qquad \text{for } (t,x) \in [0,T] \times \RR \vspace{0.1cm}
\end{equation*} 
where $\LL = r x \cdot \partial_x + \frac{1}{2} \sigma^2 x^2 \cdot \partial_{xx} $ is a differential operator. \\

The last equation has a simple interpretation. Of the two terms in the curly brackets, one will be equal to zero while the other will be negative. The first term is equal to zero when $g(t,x) > G(x)$, i.e. when the option value is greater than the intrinsic (early exercise) value, the option is not exercised early and the price function satisfies the usual Black-Scholes PDE. When the second term is equal to zero we have that $g(t,x) = G(x)$, in other words the option value is equal to the exercise value (i.e. the option is exercised). As such, the region where $g(t,x) > G(x)$ is referred to as the \textbf{continuation region} and the curve where $g(t,x) = G(x)$ is called the \textbf{exercise boundary}. Notice that it is not possible to have $g(t,x) < G(x)$ since both terms are bounded above by 0. \\

It is also worth noting that this variational inequality can be written as follows:	
\begin{equation*}
	\begin{cases}
	\partial_t g + r x \cdot \partial_x g + \frac{1}{2} \sigma^2 x^2\cdot \partial_{xx} g - r \cdot g = 0 & \qquad \left\{ (t,x): g(t,x) > G(x) \right\} 
	\\
	g(t,x) \geq G(x) & \qquad (t,x) \in [0,T] \times \RR 
	\\
	g(T,x) ~=~ G(x) &  \qquad x \in \RR
	\end{cases}
\end{equation*}
where we drop the explicit dependence on $(t,x)$ for brevity. The \textbf{free boundary} set in this problem is $F = \left\{ (t,x): g(t,x) = G(x) \right\}$ which must be determined alongside the unknown price function $g$. The set $F$ is referred to as the \textbf{exercise boundary}; once the price of the underlying asset hits the boundary, the investor's optimal action is to exercise the option immediately. \\

\subsection{The Fokker-Planck Equation}

We now turn out attention to another application of PDEs in the context of stochastic processes. Suppose we have an \Ito process on $\RR^d$ with time-independent drift and diffusion coefficients:

\[ d\vX_t = \mu(\vX_t) dt + \sigma(\vX_t) dW_t \] 
and assume that the initial point is a random vector $\vX_0$ with distribution given by a probability density function $f(\vx)$. A natural question to ask is: ``\textit{what is the probability that the process is in a given region $A \subset \RR^d$ at time $t$}?'' This quantity can be computed as an integral of the probability density function of the random vector $\vX_t$, denoted by $p(t,\vx)$:
\[ \PP\left(\vX_t \in A\right) = \int_A p(t,\vx) ~d\vx \] 
The \textbf{Fokker-Planck equation} is a partial differential equation that $p(t,\vx)$ can be shown to satisfy:
\begin{equation*}
\begin{cases}
\partial_t p(t,\vx) + \sum_{j=1}^{d} \partial_j (\mu_j(\vx) \cdot p(t,\vx)) 
\\
\qquad \quad ~~ - \frac{1}{2}  \sum_{i,j=1}^{d} \partial_{ij} (\sigma_{ij}(\vx) \cdot p(t,\vx)) = 0 \qquad \quad  (t,\vx) \in \RR_+ \times \RR^d 
\\
p(0,\vx) = f(\vx) \hspace{5.9cm} \vx \in \RR^d
\end{cases}
\end{equation*}

where $\partial_j$ and $\partial_{ij}$ are first and second order partial differentiation operators with respect to $x_j$ and $x_i$ and $x_j$, respectively. Under certain conditions on the initial distribution $f$, the above PDE admits a unique solution. Furthermore, the solution satisfies the property that $p(t,\vx)$ is positive and integrates to 1, which is required of a probability density function.  \\

As an example consider an \textbf{Ornstein-Uhlenbeck} (OU) process $X = \left(X_t\right)_{t\geq0}$ with a random starting point distributed according to an independent normal random variable with mean 0 and variance $v$. That is, $X$ satisfies the stochastic differential equation (SDE):
\[ dX_t = \kappa (\theta - X_t) ~dt + \sigma ~dW_t ~, \qquad X_0 \sim N(0,v) \] 
where $\theta$ and $\kappa$ are constants representing the mean reversion level and rate. Then the probability density function $p(t,x)$ for the location of the process at time $t$ satisfies the PDE:
\begin{equation*}
\begin{cases}
~ \partial_t p + \kappa \cdot p + \kappa(x - \theta) \cdot \partial_x p - \frac{1}{2} \sigma^2 \cdot \partial_{xx} p  = 0 \qquad \quad  (t,x) \in \RR_+ \times \RR
\\
~ p(0,x) = \frac{1}{\sqrt{2 \pi v} } \cdot e^{-\tfrac{x^2}{2 v} }
\end{cases}
\end{equation*}

Since the OU process with a fixed starting point is a Gaussian process, using a normally distributed random starting point amounts to combining the conditional distribution process with its (conjugate) prior, implying that $X_t$ is normally distributed. We omit the derivation of the exact form of $p(t,x)$ in this case.

\subsection{Stochastic Optimal Control and Optimal Stopping}
\label{sec:optimalControl}

Two classes of problems that heavily feature PDEs are \textbf{stochastic optimal control} and \textbf{optimal stopping} problems. In this section we give a brief overview of these problems along with some examples. For a thorough overview, see \cite{touzi2012optimal}, \cite{pham2009continuous} or \cite{cartea2015algorithmic}. \\

In stochastic control problems, a controller attempts to maximize a measure of success - referred to as a \textbf{performance criteria} - which depends on the path of some stochastic process by taking actions (choosing controls) that influence the dynamics of the process. In optimal stopping problems, the performance criteria depends on a stopping time chosen by the agent; the early exercise of American options discussed earlier in this chapter is an example of such a problem. \\

To discuss these in concrete terms let $X = (X_t)_{t \geq 0}$ be a controlled \Ito process satisfying the stochastic differential equation:

\[ dX^u_t = \mu(t, X_t^u, u_t) \ dt + \sigma(t, X_t^u, u_t)  \ dW_t \ , \qquad \qquad X_0^u = 0 \]  
where $u = (u_t)_{t\geq0}$ is a control process chosen by the controller from an admissible set $\As$. Notice that the drift and volatility of the process are influenced by the controller's actions. For a given control, the agent's performance criteria is:
\[ H^u(x) = \EE \biggl[ ~~ \redunderbrace{\int_{0}^{T} F(s,X_s^u,u_s) ~ds}{running reward} + \redunderbrace{\vphantom{\int_0^T} G(X_T^u)}{terminal reward}   \biggr] \]

The key to solving optimal control problems and finding the optimal control $u^*$ lies in the \textbf{dynamic programming principle} (DPP) which involves embedding the original optimization problem into a larger class of problems indexed by time, with the original problem corresponding to $t = 0$. This requires us to define:
\begin{align*}
H^u(t,x) &= \EE_{t,x} \left[ \int_{t}^{T} F(s,X_s^u,u_s) ~ds + G(X_T^u) \right]
\end{align*}
where $\EE_{t,x}[\cdot] = \EE[\cdot | X^u_t = x]$. The \textbf{value function} is the value of the performance criteria when the agent adopts the optimal control:
\[ H(t,x) = \underset{u \in \As}{\sup} ~ H^u(t,x) \]
Assuming enough regularity, the value function can be shown to satisfy a \textbf{dynamic programming equation} (DPE) also called a \textbf{Hamilton-Jacobi-Bellman} (HJB) equation. This is a PDE that can be viewed as an infinitesimal version of the DPP. The HJB equation is given by:

\begin{equation*}
\begin{cases}
\partial_t H(t,x) +  \underset{u \in \As}{\sup}  ~ \left\{ \Ls^u_t H(t,x) + F(t,x,u) \right\}  = 0
\\
H(T,x) = G(x)
\end{cases}
\end{equation*} \\
where the differential operator $\Ls^u_t$ is the \textbf{infinitesimal generator} of the controlled process $X^u$ - an analogue of derivatives for stochastic processes - given by:
\[ \LL f(t,X_t) = \underset{h \downarrow 0}{\lim} ~\frac{\EE_t[f(t+h,X_{t+h})] - f(t,X_t)}{h}  \]

\vspace{0.3cm}

Broadly speaking, the optimal control is obtained as follows:
\begin{enumerate}
	\item Solve the first order condition (inner optimization) to obtain the optimal control in terms of the derivatives of the value function, i.e. in feedback form;
	\item Substitute the optimal control back into the HJB equation, usually yielding a highly nonlinear PDE and	solve the PDE for the unknown value function;
	\item Use the value function to derive an explicit expression for the optimal control. \\
\end{enumerate}

For optimal stopping problems, the optimization problem can be written as:
\[ \underset{\tau \in \mathcal{T}}{\sup} ~ \EE \left[ G(X_\tau)  \right]  \]
where $\mathcal{T}$ is the set of admissible stopping times. Similar to the optimal control problem, we can derive a DPE for optimal stopping problem in the form of a \textbf{variational inequality} assuming sufficient regularity in the value function $H$. Namely, 
\[ \max \bigg\{ (\partial_t + \LL_t) H, ~ G-H \bigg \} = 0, \qquad \mbox{on } [0,T] \times \RR \]
The interpretation of this equation was discussed in \hyperref[sec:AmOpt]{Section \ref{sec:AmOpt}} for American-style derivatives where we discussed how the equation can be viewed as a free boundary problem. \\

It is possible to extend the problems discussed in this section in many directions by considering multidimensional processes, infinite horizons (for running rewards), incorporating jumps and combining optimal control and stopping in a single problem. This will lead to more complex forms of the corresponding dynamic programming equation. \\

Next, we discuss a number of examples of HJB equations that arise in the context of problems in quantitative finance.

\subsubsection{The Merton Problem}

In the \textbf{Merton problem}, an agent chooses the proportion of their wealth that they wish to invest in a risky asset and a risk-free asset through time. They seek to maximize the expected utility of terminal wealth at the end of their investment horizon; see \cite{merton1969lifetime} for the investment-consumption problem and \cite{merton1971optimum} for extensions in a number of directions. Once again, we assume the Black-Scholes market model: 
\begin{align*}
\frac{dS_t}{S_t} ~&=~ \mu ~dt + \sigma ~dW_t \\
\frac{dB_t}{B_t} ~&=~ r ~ dt
\end{align*}
The wealth process $X^\pi_t$ of a portfolio that invests a proportion $\pi_t$ of wealth in the risky asset and the remainder in the risk-free asset satisfies the following SDE:
\[ dX_t^\pi = \left( \pi_t (\mu - r) + r X_t^\pi \right) dt + \sigma \pi_t ~ dW_t \] 

The investor is faced with the following optimal stochastic control problem:
\[ \underset{\pi \in \As}{\sup} ~ \EE \left[ U(X_T^\pi)\right] \]
where $\As$ is the set of admissible strategies and $U(x)$ is the investor's utility function. The value function is given by:
\[ H(t,x) = \underset{\pi \in \As}{\sup} ~ \EE \left[ U(X_T^\pi) ~\middle|~ X_t^\pi = x \right] \]
which satisfies the following HJB equation:
\begin{equation*}
\begin{cases}
\partial_t H +  \underset{\pi \in \As}{\sup}  ~ \bigg\{ \big(\left( \pi (\mu -r) + rx \right) \cdot \partial_x + \tfrac{1}{2} \sigma^2 \pi^2 \partial_{xx} \big) H \bigg\}  = 0
\\
H(T,x) = U(x)
\end{cases}
\end{equation*}
If we assume an exponential utility function with risk preference parameter $\gamma$, that is $U(x) = -e^{-\gamma x}$, then the value function and the optimal control can be obtained in closed-form: 
\begin{align*}
H(t,x)  &= - \exp \left[ {-x \gamma e^{r(T-t)} - \tfrac{\lambda^2}{2}  (T-t)	} \right] 
\\
\pi^*_t &= \frac{\lambda}{\gamma \sigma} e^{-r(T-t)}
\end{align*}
where $\lambda = \frac{\mu - r}{\sigma}$ is the market price of risk. \\

It is also worthwhile to note that the solution to the Merton problem plays an important role in the \textbf{substitute hedging} and \textbf{indifference pricing} literature, see e.g. \cite{henderson2002substitute} and \cite{henderson2004utility}.

\subsubsection{Optimal Execution with Price Impact}

Stochastic optimal control, and hence PDEs in the form of HJB equations, feature prominently in the algorithmic trading literature, such as in the classical work of \cite{almgren2001optimal} and more recently \cite{cartea2015optimal} and \cite{cartea2016incorporating} to name a few. Here we discuss a simple algorithmic trading problem with an investor that wishes to liquidate an inventory of shares but is subject to price impact effects when trading too quickly. The challenge then involves balancing this effect with the possibility of experiencing a negative market move when trading too slowly. \\ 

We begin by describing the dynamics of the main processes underlying the model. The agent can control their (liquidation) \textbf{trading rate} $\nu_t$ which in turn affects their \textbf{inventory level} $Q^\nu_t$ via:
\[ dQ^\nu_t = -\nu_t ~dt, \qquad Q_0^\nu = q \]
Note that negative values of $\nu$ indicate that the agent is buying shares. The price of the underlying asset $S_t$ is modeled as a Brownian motion that experiences a \textbf{permanent price impact} due to the agent's trading activity in the form of a linear increase in the drift term:
\[ dS_t^\nu = - b \nu_t ~dt + \sigma ~dW_t, \qquad S_0^\nu = S \]
By selling too quickly the agent applies increasing downward pressure (linearly with factor $b>0$) on the asset price which is unfavorable to a liquidating agent. Furthermore, placing larger orders also comes at the cost of increased \textbf{temporary price impact}. This is modeled by noting that the cashflow from a particular transaction is based on the \textbf{execution price} $\widehat{S}_t$ which is linearly related to the fundamental price (with a factor of $k>0$):
\[ \widehat{S}_t^\nu = S_t^\nu - k \nu_t \]
The \textbf{cash process} $X_t^\nu$ evolves according to:
\[ dX_t^\nu = \widehat{S}_t^\nu \nu_t ~dt, \qquad X^\nu_0 = x \] 

With the model in place we can consider the agent's performance criteria, which consists of maximizing their terminal cash and penalties for excess inventory levels both at the terminal date and throughout the liquidation horizon. The performance criteria is
\[ H^\nu(t,x,S,q) = \EE_{t,x,S,q} \bigg[ \redunderbrace{\vphantom{\int_t^T} X_T^\nu}{\parbox{1.5cm}{\centering terminal \\ cash}} + ~~ \redunderbrace{\vphantom{\int_t^T} Q_T^\nu \left(S_T^\nu - \alpha Q_T^\nu \right)}{\parbox{1.5cm}{\centering terminal inventory}} - \redunderbrace{\phi \int_t^T \left( Q^\nu_u \right)^2 du}{running inventory} ~~ \bigg] \] 
where $\alpha$ and $\phi$ are preference parameters that control the level of penalty for the terminal and running inventories respectively. The value function satisfies the HJB equation:

\begin{equation*}
\begin{cases}
(\partial_t + \tfrac{1}{2} \sigma^2 \partial_{SS}) H - \phi q^2 
\\
\qquad + ~ \underset{\nu}{\sup}  \left\{ \left(\nu(S - k\nu)\partial_x - b\nu \cdot \partial _S - \nu \partial_q \right) H \right\} = 0
\\
H(t,x,S,q) = x + Sq - \alpha q^2
\end{cases}
\end{equation*}

Using a carefully chosen ansatz we can solve for the value function and optimal control:
\begin{align*}
H(t,x,S,q) &= x + qS + \left( h(t) - \tfrac{b}{2} \right) q^2
\\
\nu_t^* &= \gamma \cdot \frac{\zeta e^{\gamma(T-t)} + e^{-\gamma(T-t)}}{\zeta e^{\gamma(T-t)} - e^{-\gamma(T-t)}} \cdot Q_t^{\nu^*}
\\
\mbox{where } \quad h(t) &= \sqrt{k \phi} \cdot \frac{1 + \zeta e^{2\gamma(T-t)}}{1 - \zeta e^{2\gamma(T-t)}},
\qquad 
\gamma = \sqrt{\frac{\phi}{k}} ,
\qquad
\zeta = \frac{\alpha - \tfrac{1}{2}b + \sqrt{k \phi}}{\alpha - \tfrac{1}{2}b - \sqrt{k \phi}}
\end{align*}

For other optimal execution problems the interested reader is referred to Chapter 6 of \cite{cartea2015algorithmic}. 
 
\subsubsection{Systemic Risk}

Yet another application of PDEs in optimal control is the topic of \cite{carmona2015systemic}. The focus in that paper is on \textbf{systemic risk} - the study of instability in the entire market rather than a single entity - where a number of banks are borrowing and lending with the central bank with the target of being at or around the average monetary reserve level across the economy. Once a characterization of optimal behavior is obtained, questions surrounding the stability of the system and the possibility of multiple defaults can be addressed. This is an example of a \textbf{stochastic game}, with multiple players determining their preferred course of action based on the actions of others. The object in stochastic games is usually the determination of \textbf{Nash equilibria} or sets of strategies where no player has an incentive to change their action. \\

The main processes underlying this problem are the log-monetary reserves of each bank denoted $X^i = \left( X^i_t \right)_{t \geq 0}$ and assumed to satisfy the SDE:
\[ dX_t^i = \left[a \left( \overline{X}_t - X^i_t \right) + \alpha_t^i \right] dt + \sigma ~ d\widetilde{W}_t^i \]
where $\widetilde{W}^i_t = \rho W_t^0 + \sqrt{1-\rho^2} W_t^i$ are Brownian motions correlated through a common noise process, $\overline{X}_t$ is the average log-reserve level and $\alpha_t^i$ is the rate at which bank $i$ borrows from or lends to the central bank.
The interdependence of reserves appears in a number of places: first, the drift contains a mean reversion term that draws each bank's reserve level to the average with a mean reversion rate $a$; second, the noise terms are driven partially by a common noise process. \\

The agent's control in this problem is the borrowing/lending rate $\alpha^i$. Their aim is to remain close to the average reserve level at all times over some fixed horizon. Thus, they penalize any deviations from this (stochastic) average level in the interim and at the end of the horizon. They also penalize borrowing and lending from the central bank at high rates as well as borrowing (resp. lending) when their own reserve level is above (resp. below) the average level. Formally, the performance criterion is given by:
\[ J^i \left(\alpha^1,...,\alpha^N\right) =  \EE \left[ \int_0^T f_i \left( \vX_t, \alpha^i_t \right) dt + g_i \left( X_T^i \right) \right] \] 
where the running penalties are:
\[ f_i(\vx, \alpha^i) = \redunderbrace{\tfrac{1}{2} \left( \alpha^i \right)^2}{\tiny \centering  \parbox{1.75cm}{\centering excessive lending \\ or borrowing}} - \redunderbrace{q \alpha^i \left(\overline{x} - x^i \right)}{ \tiny \parbox{2.2cm}{\centering borrowing/lending in \\ ``the wrong direction''}} + \redunderbrace{\tfrac{\epsilon}{2}  \left(\overline{x} - x^i \right)^2}{\centering  \tiny \parbox{1.75cm}{\centering deviation from the average level}}  \]
and the terminal penalty is:
\[ g_i(\vx) =  \redunderbrace{\tfrac{c}{2}  \left(\overline{x} - x^i \right)^2}{\centering  \tiny \parbox{1.75cm}{\centering deviation from the average level}} \]

where $c,q,$ and $\epsilon$ represent the investor's preferences with respect to the various penalties. Notice that the performance criteria for each agent depends on the strategies and reserve levels of all the agents including themselves. Although the paper discusses multiple approaches to solving the problem (Pontryagin stochastic maximum principle and an alternative forward-backward SDE approach), we focus on the HJB approach as this leads to a system of nonlinear PDEs. Using the dynamic programming principle, the HJB equation for agent $i$ is:

\begin{equation*}
\begin{cases}
{\displaystyle
\partial_t V^i + \underset{\alpha^i}{\inf} ~ \bigg\{ \sum_{j=1}^{N} \left[a(\overline{x} - x^j) + \alpha^j\right] \partial_j V^i }
\\
{ \displaystyle \hspace{2.5cm} + \frac{\sigma^2}{2} \sum_{j,k=1}^{N} \left( \rho^2 + \delta_{jk} (1-\rho^2) \right) \partial_{jk} V^i }
\\
{ \displaystyle \hspace{2.5cm} + \tfrac{(\alpha^i)^2}{2} - q \alpha^i (\overline{x} - x^i) + \tfrac{\epsilon}{2}  \left(\overline{x} - x^i \right)^2 \bigg\} = 0 }
\\
V^i(T,\vx) = \tfrac{c}{2}  \left(\overline{x} - x^i \right)^2
\end{cases}
\end{equation*}
\vspace{0.7cm}

Remarkably, this system of PDEs can be solved in closed-form to obtain the value function and the optimal control for each agent:
\begin{align*}
V^i(t,\vx) &= \frac{\eta(t)}{2} \left(\overline{x} - x^i \right)^2 + \mu(t) 
\\
\alpha_t^{i,*} &=  \bigg( q + \left(1 - \tfrac{1}{N} \right) \cdot \eta(t) \bigg) \left( \overline{X}_t - X^i_t \right)
\\
\mbox{where } \qquad  \eta(t) &= \frac{ -(\epsilon - q)^2 \left( e^{(\delta^+ - \delta^-)(T-t)} - 1 \right) - c \left( \delta^+ e^{(\delta^+ - \delta^-)(T-t)} - \delta^- \right)}{ \left( \delta^-e^{(\delta^+ - \delta^-)(T-t)} - \delta^+ \right) - c (1 - \tfrac{1}{N^2}) \left( e^{(\delta^+ - \delta^-)(T-t)} - 1 \right)}
\\ \mu(t) &= \tfrac{1}{2} \sigma^2 (1-\rho^2) \left( 1 - \tfrac{1}{N}  \right) \int_t^T \eta(s) ~ds
\\ 
\delta^{\pm} &= - (a+q) \pm \sqrt{R}, 
\qquad \qquad 
R = (a+q)^2 + \Big( 1 - \tfrac{1}{N^2} \Big)(\epsilon - q^2) \\
\end{align*}

\clearpage

\subsection{Mean Field Games}

The final application of PDEs that we will consider is that of \textbf{mean field games} (MFGs). In financial contexts, MFGs are concerned with modeling the behavior of a large number of small interacting market participants. In a sense, it can be viewed as a limiting form of the Nash equilibria for finite-player stochastic game (such as the interbank borrowing/lending problem from the previous section) as the number of participants tends to infinity. Though it may appear that this would make the problem more complicated, it is often the case that this simplifies the underlying control problem. This is because in MFGs, agents need not concern themselves with the actions of every other agent, but rather they pay attention only to the aggregate behavior of the other agents (the mean field). It is also possible in some cases to use the limiting solution to obtain approximations for Nash equilibria of finite player games when direct computation of this quantity is infeasible. The term `'mean field'' originates from mean field theory in physics which, similar to the financial context, studies systems composed of large numbers of particles where individual particles have negligible impact on the system. A mean field game typically consists of:

\begin{enumerate}
	\item An \textbf{HJB equation} describing the optimal control problem of an individual; 
	\item A \textbf{Fokker-Planck equation} which governs the dynamics of the aggregate behavior of all agents. \\
\end{enumerate} 

Much of the pioneering work in MFGs is due to \cite{huang2006} and \cite{lasry2007mean}, but the focus of our exposition will be on a more recent work by \cite{cardaliaguet2017mean}. Building on the optimal execution problem discussed earlier in this chapter, \cite{cardaliaguet2017mean} propose extensions in a number of directions. First, traders are assumed to be part of a mean field game and the price of the underlying asset is impacted permanently, not only by the actions of the agent, but by the aggregate behavior of all agents acting in an optimal manner. In addition to this aggregate permanent impact, an individual trader faces the usual temporary impact effects of trading too quickly. The other extension is to allow for varying preferences among the traders in the economy. That is, traders may have different tolerance levels for the size of their inventories both throughout the investment horizon and at its end. Intuitively, this framework can be thought of as the agents attempting to ``trade optimally within the crowd.'' \\

Proceeding to the mathematical description of the problem, we have the following dynamics for the various agents' \textbf{inventory and cash processes} (indexed by a superscript $a$):   

\begin{align*}
dQ^a_t &= \nu^a_t ~dt, && Q_0^a = q^a
\\
dX_t^a &= -\nu^a_t \left( S_t + k \nu_t^a \right) ~dt,  && X^a_0 = x^a
\end{align*}

An important deviation from the previous case is the fact that the permanent price impact is due to the \textbf{net sum of the trading rates of all agents}, denoted by $\mu_t$:
\[ dS_t = \kappa \mu_t ~ dt + \sigma ~dW_t \]

Also, the value function associated with the optimal control problem for agent $a$ is given by:
\[ H^a(t,x,S,q) = \underset{\nu}{\sup} ~ \EE_{t,x,S,q} \bigg[ \redunderbrace{ \vphantom{\int_t^T} X_T^a}{\parbox{1.3cm}{\centering terminal \\ cash}} + ~~ \redunderbrace{\vphantom{\int_t^T} Q_T^a \left(S_T - \alpha^a Q_T^a \right)}{\parbox{1.5cm}{\centering terminal inventory}} ~-~ \redunderbrace{\phi^a \int_t^T \left( Q^a_u \right)^2 du}{running inventory} ~~ \bigg] \] 

Notice that each agent $a$ has a \textit{different} value of $\alpha^a$ and $\phi^a$ demonstrating their differing preferences. As a consequence, an agent can be represented by their preferences $a = (\alpha^a,\phi^a)$. The HJB equation associated with the agents' control problem is:
\begin{equation*}
\begin{cases}
\left( \partial_t + \tfrac{1}{2} \sigma^2 \partial_{SS} \right) H^a - \phi^a q^2 + \kappa \mu \cdot \partial_S H^a +  \underset{\nu}{\sup} ~ \bigg\{ \left(  \nu \cdot \partial_q - \nu (S + k\nu) \cdot \partial_x \right) H^a \bigg\} = 0
\\
H^a(T,x,S,q;\mu) = x + q (S - \alpha^a q)
\end{cases}
\end{equation*}
This can be simplified using an ansatz to:
\begin{equation*}
\begin{cases}
-\kappa \mu q = \partial_t h^a - \phi^a q^2 + \underset{\nu}{\sup} ~ \bigg\{ \nu \cdot \partial_q h^a - k\nu^2  \bigg\} 
\\
h^a(T,q) = - \alpha^a q^2
\end{cases}
\end{equation*}
Notice that the PDE above requires agents to know the net trading flow of the mean field $\mu$, but that this quantity itself depends on the value function of each agent which we have yet to solve for. To resolve this issue we first write the optimal control of each agent in feedback form:
\[ \nu^a(t,q) = \frac{\partial_q h^a(t,q)}{2k} \]

Next, we assume that the distribution of inventories and preferences of agents is captured by a density function $m(t,dq,da)$. With this, the net flow $\mu_t$ is simply given by the aggregation of all agents' optimal actions:
\[ \mu_t = \int_{(q,a)} \redunderbrace{\frac{\partial h^a(t,q)}{2k}}{\tiny \parbox{2cm}{\centering trading rate of agent \\ with inventory $q$ \\ and preferences $a$}} ~\redunderbrace{\vphantom{\frac{\partial h^a(t,q)}{2k}} m(t,dq,da)}{\tiny \parbox{2.4cm}{\centering aggregated according to \\ distribution of agents}}  \] 
In order to compute the quantity at different points in time we need to understand the evolution of the density $m$ through time. This is just an application of the Fokker-Planck equation, as $m$ is a density that depends on a stochastic process (the inventory level). If we assume that the initial density of inventories and preferences is $m_0(q,a)$, we can write the Fokker-Planck equation as:
\begin{equation*}
\begin{cases}
\partial_t m + \partial_q  \bigg( m \cdot \redunderbrace{\frac{\partial h^a(t,q)}{2k}}{\tiny \parbox{1.75cm}{\centering drift of inventory \\ process $Q^a_t$ under \\ optimal controls}} \bigg) =0
\\
m(0,q,a) = m_0(q,a)
\end{cases}
\end{equation*} 
  
The full system for the MFG in the problem of \cite{cardaliaguet2017mean} involves the combined HJB and Fokker-Planck equations with the appropriate initial and terminal conditions:
\begin{equation*}
\begin{cases}
~~ \displaystyle -\kappa \mu q = \partial_t h^a - \phi^a q^2 + \frac{\left( \partial_q h^a \right)^2}{4k} & \qquad \qquad \text{\color{red} (HJB equation - optimality)}
\\
~~ \displaystyle H^a(T,x,S,q;\mu) = x + q (S - \alpha^a q) & \qquad \qquad \text{\color{red} (HJB terminal condition)}
\\
~
\\
~~ \displaystyle \partial_t m + \partial_q  \bigg( m \cdot \frac{\partial h^a(t,q)}{2k} \bigg) = 0 & \qquad \qquad \text{\color{red} (FP equation - density flow)}
\\
~~ \displaystyle m(0,q,a) = m_0(q,a) & \qquad \qquad \text{\color{red} (FP initial condition)}
\\
~
\\
~~ \displaystyle \mu_t = \int_{(q,a)} \frac{\partial h^a(t,q)}{2k} ~m(t,dq,da) & \qquad \qquad \text{\color{red} (net trading flow)}
\end{cases}
\end{equation*} \vspace{0.3cm}

Assuming identical preferences $\alpha^a = \alpha, \phi^a = \phi$ allows us to find a closed-form solution to this PDE system. The form of the solution is fairly involved so we refer the interested reader to the details in \cite{cardaliaguet2017mean}.


%% file: NumericalMethods.tex
\chapter{Numerical Methods for PDEs} \label{chap:numerical}
\setcounter{section}{3}

Although it is possible to obtain closed-form solutions to PDEs, more often we must resort to numerical methods for arriving at a solution. In this chapter we discuss some of the approaches taken to solve PDEs numerically. We also touch on some of the difficulties that may arise in these approaches involving stability and computational cost, especially in higher dimensions. This is by no means a comprehensive overview of the topic to which a vast amount of literature is dedicated. Further details can be found in \cite{burden2001numerical}, \cite{achdou2005computational} and \cite{brandimarte2013numerical}. \\

\subsection{Finite Difference Method}

It is often the case that differential equations cannot be solved
analytically, so one must resort to numerical methods to solve them.
One of the most popular numerical methods is the \textbf{finite difference
method}. As its name suggests, the main idea behind this method is
to approximate the differential operators with difference operators and apply them to a discretized version of the unknown function in the differential equation.

\subsubsection{Euler's Method }

Arguably, the simplest finite difference method is \textbf{Euler's method}
for \textbf{ordinary differential equations} (ODEs). Suppose we have the following initial value problem
\begin{equation*}
\begin{cases}
y'(t)=f(t)
\\
y(0)=y_0
\end{cases}
\end{equation*}
for which we are trying to solve for the function $y(t)$. By the Taylor series expansion, we can write 
\[
y(t+h)=y(t)+\frac{y'(t)}{1!} \cdot h+\frac{y''(t)}{2!} \cdot h^{2}+\cdots
\]
for any infinitely differentiable real-valued function $y$. If $h$ is small enough, and if the derivatives of $y$ satisfy some regularity conditions, then terms of order $h^{2}$ and higher are negligible and we can make the approximation
\[
y(t+h)\approx y(t)+y'(t)\cdot h
\]
As a side note, notice that we can rewrite this equation as 
\[
y'(t)\approx\frac{y(t+h)-y(t)}{h}
\]
which closely resembles the definition of a derivative;
\[
y'(t)=\lim_{h\rightarrow0}\frac{y(t+h)-y(t)}{h}.
\]
Returning to the original problem, note that we know the exact value of $y'(t)$, namely $f(t)$, so that we can write 
\[
y(t+h)\approx y(t)+f(t)\cdot h.
\]

At this point, it is helpful to introduce the notation for the discretization
scheme typically used for finite difference methods. Let $\left\{ t_{i}\right\} $
be the sequence of values assumed by the time variable, such that
$t_{0}=0$ and $t_{i+1}=t_{i}+h$, and let $\left\{ y_{i}\right\} $
be the sequence of approximations of $y(t)$ such that
$y_{i}\approx y(t_{i})$. The expression above can be rewritten
as
\[
y_{i+1}\approx y_{i}+f(t_{i})\cdot h,
\]
which allows us to find an approximation for the value of $y\left(t_{i+1}\right)\approx y_{i+1}$
given the value of $y_{i}\approx y(t_{i})$. Using Euler's method, we can find numerical approximations for $y(t)$ for any value of $t>0$. \\

\subsubsection{Explicit versus implicit schemes}

In the previous section, we developed Euler's method for a simple
initial value problem. Suppose one has the slightly different problem where the source term $f$ is now a function of both $t$ and $y$.

\begin{equation*}
\begin{cases}
y'(t)=f(t,y)
\\
y(0)=y_{0}
\end{cases}
\end{equation*}
A similar argument
as before will now lead us to the expression for $y_{i+1}$
\[
y_{i+1}\approx y_{i}+f(t_{i},y_{i})\cdot h,
\]
where $y_{i+1}$ is \textit{explicitly} written as a sum of terms that depend
only on time $t_{i}$. Schemes such as this are called \textbf{explicit}. Had
we used the approximation
\[
y(t-h)\approx y(t)-y'(t)\cdot h
\]
instead, we would arrive at the slightly different expression for
$y_{i+1}$
\[
y_{i+1}\approx y_{i}+f(t_{i+1},y_{i+1})\cdot h,
\]
where the term $y_{i+1}$ appears in both sides of the equation and
no explicit formula for $y_{i+1}$ is possible in general. Schemes
such as this are called \textbf{implicit}. In the general case, each step in
time in an implicit method requires solving the expression above for
$y_{i+1}$ using a root finding technique such as \textbf{Newton's method}
or other fixed point iteration methods. \\

Despite being easier to compute, explicit methods are generally known
to be numerically unstable for a large range of equations (especially so-called
\textbf{stiff problems}), making them unusable for most practical situations.
Implicit methods, on the other hand, are typically both more computationally
intensive and more numerically stable, which makes them more commonly
used. An important measure of numerical stability for finite difference
methods is \textbf{A-stability}, where one tests the stability of the method
for the (linear) test equation $y'(t,y)=\lambda y(t)$,
with $\lambda<0$. While the implicit Euler method is stable for all
values of $h>0$ and $\lambda<0$, the explicit Euler method is stable
only if $\left|1+h\lambda\right|<1$, which may require using a small
value for $h$ if the absolute value of $\lambda$ is high. Of course,
all other things being equal, a small value for $h$ is undesirable
since it means a finer grid is required, which in turn makes the numerical method
more computationally expensive.

\subsubsection{Finite difference methods for PDEs}

In the previous section, we focused our discussion on methods for
numerically solving ODEs. However, finite difference methods can be
used to solve PDEs as well and the concepts presented above can also
be applied in PDEs solving methods. Consider the boundary problem
for the heat equation in one spatial dimension, which describes the dynamics of heat transfer in a rod of length $l$:

\begin{equation*}
\begin{cases}
\partial_t u =\alpha^{2} \cdot \partial_{xx} u
\\
u(0,x)=u_{0}(x) 
\\
u(t,0)=u(t,l)=0
\end{cases}
\end{equation*}

We could approximate the differential operators in the equation above
using a \textbf{forward difference} operator for the partial derivative in
time and a \textbf{second-order central difference} operator for the partial
derivative in space. Using the notation $u_{i,j}\approx u(t_{i},x_{j})$,
with $t_{i+1}=t_{i}+k$ and $x_{j+1}=x_{j}+h$, we can rewrite the
equation above as a system of linear equations
\[
\frac{u_{i+1,j}-u_{i,j}}{k}=\alpha^{2}\left(\frac{u_{i,j-1}-2u_{i,j}+u_{i,j+1}}{h^{2}}\right),
\]
where $i=1,2,\dots,N$ and $j=1,2,\dots,N$, assuming we are using
the same number of discrete points on both dimensions. In this two
dimensional example, the points $\left(t_{i},x_{j}\right)$ form a
two dimensional grid of size $\mathcal{O}\left( N^{2}\right)$. For a $d$-dimensional
problem, a $d$-dimensional grid with
size $\mathcal{O}\left(N^{d} \right)$ would be required. In practice, the exponential growth of the grid in the number of dimensions rapidly makes the method unmanageable,
even for $d = 4$. \textit{This is an important shortcoming of finite difference methods in general}. \\

\begin{figure}[h!]
	\centering
	\begin{tikzpicture}   
	
		\draw[step=0.5cm,black!30,very thin] (-2,-2) grid (2,2);
	
	    \draw[red,thick] (-2.0,2.0) -- (-2.0,-2.0);
	    \draw[blue,thick] (-2.0,-2.0) -- (2.0,-2.0);
	
	    \draw[cadmiumgreen,thick] (0,0) -- (0,-0.5);
	    \draw[cadmiumgreen,thick] (0,0) -- (-0.5,0);
	   	\draw[cadmiumgreen,thick] (0,0) -- (0.5,0);
	
	
	    \foreach \x in {-2.0,-1.5,...,2.0}{
	    \foreach \y in {-2.0,-1.5,...,2.0}{
	    \fill[black!80] (\x,\y) circle[radius=1.3pt];
	    }}
	    
	    \fill[cadmiumgreen] (0,0) circle[radius=1.3pt];
	    
	    \node at (-2.5,0) {\small $x$};    
	    \node at (0,-2.5) {\small $t$};
	    
	    \node at (0,0.25) {\color{cadmiumgreen} \scriptsize $(t_i,x_j)$};
		
	    \node[label={[align=center,label distance=0cm,text depth=-0.8cm]: \color{blue} \scriptsize boundary }] at (1.875,-2.1) {}; 
	
	    \node[label={[align=center,label distance=0cm,text depth=-0.8cm]: \color{blue}\color{blue} \scriptsize condition}] at (1.875,-2.37) {};
	    
	    \node[label={[align=center,label distance=0cm,text depth=0,rotate=90]: \color{red} \scriptsize initial}] at (-2.37,1.875) {};
	
	    \node[label={[align=center,label distance=0cm,text depth=0,rotate=90]: \color{red} \scriptsize condition}] at (-2.1,1.875) {};
	    
	    \node at (0,2.5) {\footnotesize mesh grid points};  
	    
    	\node at (-0.5,-3.25) {~~~};  
		
	\end{tikzpicture} \qquad \begin{tikzpicture}   
		
			\draw[rectangle,fill=red!10] (-2,-2,-2) -- (-2,-2,2) -- (-2,2,2) -- (-2,2,-2);		
			\draw[rectangle,fill=blue!10] (-2,-2,-2) -- (-2,-2,2) -- (2,-2,2) -- (2,-2,-2);	
						
 			\foreach \x in {-2.0,0.0,2.0}{
			\foreach \y in {-2.0,0.0,2.0}{
	        \foreach \z in {-2.0,0.0,2.0}{
	              \draw[black!30,very thin] (\x,-2,\z) -- (\x,2,\z);
	              \draw[black!30,very thin] (\x,\y,-2) -- (\x,\y,2);
				  \draw[black!30,very thin] (-2,\y,\z) -- (2,\y,\z);
	        }}}
		
		    \draw[cadmiumgreen,thick] (0,0,0) -- (-2,0,0);
		    \draw[cadmiumgreen,thick] (0,0,0) -- (2,0,0);
		   	\draw[cadmiumgreen,thick] (0,0,0) -- (0,0,-2);
		   	\draw[cadmiumgreen,thick] (0,0,0) -- (0,2,0);
		
		    \foreach \x in {-2.0,0.0,...,2.0}{
		    \foreach \y in {-2.0,0.0,...,2.0}{
   		    \foreach \z in {-2.0,0.0,...,2.0}{
		    \fill[black!80] (\x,\y,\z) circle[radius=1.7pt];		    
		    }}}
		    
      	    \node at (-3.5,-0.5) {\small $x$};    
      	    \node at (3,-2.5) {\small $y$};
      	    \node at (-0.5,-3.5) {\small $t$};
			\node at (0.1,-0.3) {\color{cadmiumgreen} \small $(t_i,x_j,y_k)$};   
      	    
			\fill[cadmiumgreen] (0,0,0) circle[radius=1.7pt];      	    
		    
		    
		    
			
		    \node[label={[align=center,label distance=0cm,text depth=-0.8cm]: \color{blue} \scriptsize boundary }] at (1.5,-2.8) {}; 
		
		    \node[label={[align=center,label distance=0cm,text depth=-0.8cm]: \color{blue}\color{blue} \scriptsize condition}] at (1.5,-3.05) {};
		    
		    \node[label={[align=center,label distance=0cm,text depth=0]: \color{red} \scriptsize initial}] at (-3,1.875) {};
		
		    \node[label={[align=center,label distance=0cm,text depth=0]: \color{red} \scriptsize condition}] at (-3,1.625) {};
		    
		    \node at (0.6,3.1) {\footnotesize mesh grid points};    
			
		\end{tikzpicture}
	
	\caption{Illustration of finite difference methods for solving PDEs in two (left) and three (right) dimensions. The known function value on the boundaries is combined with finite differences to solve for the value of function on a grid in the interior of the region where it is defined.}
	\label{fig:finiteDiff}
\end{figure}
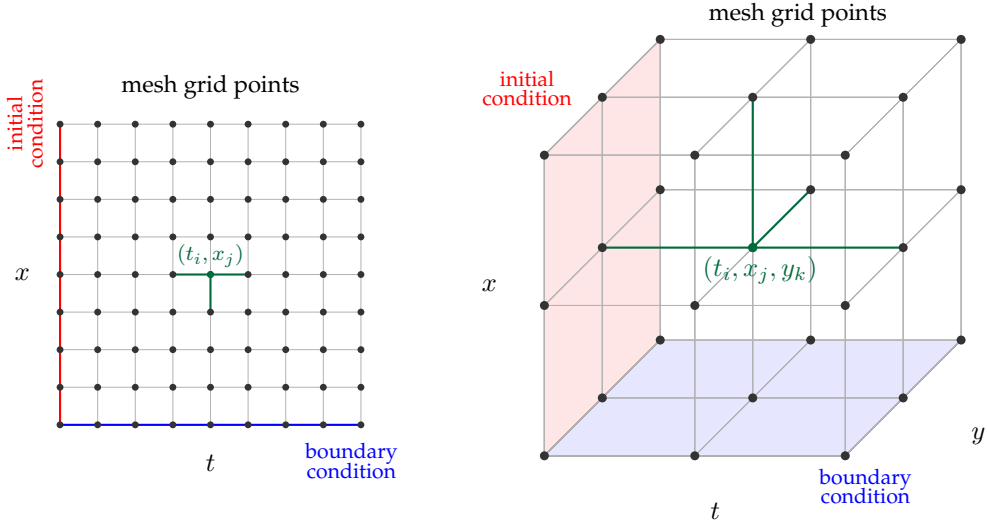

The scheme developed above is known as the \textbf{forward difference} method
or \textbf{FTCS (forward in time, central in space)}. It is easy to verify
that this scheme is explicit in time, since we can write the $u_{i+1,\cdot}$
terms as a linear combination of previously computed $u_{i,\cdot}$
terms. The number of operations necessary to advance each step in
time with this method should be $\mathcal{O}\left(N^{2}\right)$. Unfortunately,
this scheme is also known to be unstable if $h$ and $k$ do not satisfy
the inequality $\alpha^{2}\frac{k}{h^{2}}\leq\frac{1}{2}$. \\

Alternatively, we could apply the \textbf{Backward Difference} method or \textbf{BTCS
(backward in time, central in space)} using the following equations:

\[
\frac{u_{i+1,j}-u_{i,j}}{k}=\alpha^{2}\left(\frac{u_{i+1,j-1}-2u_{i+1,j}+u_{i+1,j+1}}{h^{2}}\right).
\]
This scheme is implicit in time since it is not possible to write
the $u_{i+1,\cdot}$ terms as a function of just the previously computed
$u_{i,\cdot}$ terms. In fact, each step in time requires solving system of linear equations of size $\mathcal{O}\left(N^{2}\right)$. The number of operations
necessary to advance each step in time with this method is
$\mathcal{O}\left(N^{3}\right)$ when using methods such as Gaussian elimination
to solve the linear system. On the other hand, this scheme is also
known to be \textit{unconditionally stable}, independently of the values for
$h$ and $k$.

\subsubsection{Higher order methods}

All numerical methods for solving PDEs have errors due to many sources
of inaccuracies. For instance, \textbf{rounding error} is related to the
floating point representation of real numbers. Another important category
of error is \textbf{truncation error}, which can be understood as the error
due to the Taylor series expansion truncation. Finite difference methods
are usually classified by their respective truncation errors. \\

All finite methods discussed so far are low order methods. For instance,
the Euler's methods (both explicit and implicit varieties) are \textbf{first-order
methods}, which means that the global truncation error is proportional
to $h$, the discretization granularity. However,
a number of alternative methods have lower truncation errors.
For example, the \textbf{Runge-Kutta 4th-order method}, with a global truncation
error proportional to $h^{4}$, is widely used, being the most known
method of a family of finite difference methods, which cover even
14th-order methods. Many Runge-Kutta methods are specially suited
for solving stiff problems. \\

\subsection{Galerkin methods}

In finite difference methods, we approximate the continuous differential
operator by a discrete difference operator in order to obtain a numerical
approximation of the function that satisfies the PDE. The function's
domain (or a portion of it) must also be discretized so that numerical
approximations for the value of the solution can be computed at the
points of the so defined spatial-temporal grid. Furthermore, the value
of the function on off-grid points can also be approximated by techniques
such as interpolation. \\

\textbf{Galerkin methods} take an alternative approach: given a finite
set of basis functions on the same domain, the goal is to find a linear
combination of the basis functions that approximates the solution
of the PDE on the domain of interest. This problem translates
into a variational problem where one is trying to find maxima or minima
of functionals. \\

More precisely, suppose we are trying to solve the equation $F(x)=y$
for $x$, where $x$ and $y$ are members of spaces of functions $X$ and $Y$ respectively and that $F:X\rightarrow Y$ is a (possibly
non-linear) functional. Suppose also that $\left\{ \phi_{i}\right\} _{i=1}^{\infty}$
and $\left\{ \psi_{j}\right\} _{j=1}^{\infty}$ form linearly independent
bases for $X$ and $Y$. According to the Galerkin method,
an approximation for $x$ could be given by
\[
x_{n}=\sum_{i=1}^{n}\alpha_{i}\phi_{i}
\]
where the $\alpha_{i}$ coefficients satisfy the equations
\[
\left\langle F\left(\sum_{i=1}^{n}\alpha_{i}\phi_{i}\right),\psi_{j}\right\rangle =\left\langle y,\psi_{j}\right\rangle ,
\]
for $j=1,2,\dots,n$.\footnote{\url{https://www.encyclopediaofmath.org/index.php/Galerkin\_method }}
Since the inner products above usually involve non-trivial integrals,
one should carefully choose the bases to ensure the equations are more manageable.\\

\subsection{Finite Element Methods}

\textbf{Finite element methods} can be understood as a special case of 
Galerkin methods. Notice that in the general case presented above,
the approximation $x_{n}$ may not be well-posed, in the sense that
the system of equations for $\alpha_{i}$ may have no solution or
it may have multiple solutions depending on the value of $n$. Additionally,
depending on the choice of $\phi_{i}$ and $\psi_{j}$, $x_{n}$ may
not converge to $x$ as $n\rightarrow\infty$. Nevertheless, one could
discretize the domain in small enough regions (called elements) so
that the approximation is locally satisfactory in each region. Adding
boundary consistency constraints for each region intersection (as
well as for the outer boundary conditions given by the problem definition)
and solving for the whole domain of interest, one can come up with
a globally fair numerical approximation for the solution to the PDE. \\

In practice, the domain is typically divided in triangles or quadrilaterals
(two-dimensional case), tetrahedra (three-dimensional case) or more
general geometrical shapes in higher dimensions in a process known
as \textbf{triangulation}. Typical choices for $\phi_{i}$ and $\psi_{j}$
are such that the inner product equations above reduce to a system of
algebraic equations for steady state problems or a system of ODEs in the case
of time-dependent problems. If the PDE is linear, those systems will
be linear as well, and they can be solved using methods such as Gaussian
elimination or iterative methods such as Jacobi or Gauss-Seidel. If
the PDE is not linear, one may need to solve systems of nonlinear
equations, which are generally more computationally expensive. One of the major advantages of the finite element methods over finite difference methods, is that finite elements can effortlessly handle complex boundary geometries, which typically arise in physical
or engineering problems, whereas this may be very difficult to achieve with finite difference
algorithms. \\

\subsection{Monte Carlo Methods}

One of the more fascinating aspects of PDEs is how they are intimately related to stochastic processes. This is best exemplified by the \textbf{Feynman-Kac theorem}, which can be viewed in two ways:
\begin{itemize}
	\item It provides a solution to a certain class of linear PDEs, written in terms of an expectation involving a related stochastic process;
	\item It gives a means by which certain expectations can be computed by solving an associated PDE. 
\end{itemize}  
For our purposes, we are interested in the first of these two perspectives. \\

The theorem is stated as follows: the solution to the partial differential equation 
\[
\begin{cases}
\partial_t h + a(t,x) \cdot \partial_x h + \frac{1}{2} b(t,x)^2 \cdot \partial_{xx} h + g(t,x) \cdot h(t,x) ~=~  c(t,x) \cdot h(t,x) \\
h(T,x) ~=~ H(x)
\end{cases}
\]
admits a stochastic representation given by
\[ h(t,x) ~=~ \EE^{\PP^*}_{t,x} \left[ \int_t^T e^{-\int_t^u c(s,X_s) ~ds} \cdot g(u,X_u) ~du + H(X_T) \cdot e^{-\int_t^T c(s,X_s) ~ds} \right] \]
where $\EE_{t,x}[~\cdot~] = \EE\left[ ~\cdot~ \middle| X_t = x \right] $ and the process $X = (X_t)_{t \geq 0}$ satisfies the SDE:
\[ dX_t ~=~ a(t,X_t) ~ dt + b(t,X_t) ~ dW_t^{\PP^*} \] 
where $W^{\PP^*} = \left( W_t^{\PP^*} \right)_{t \geq 0}$ is a standard Brownian motion under the probability measure $\PP^*$. This representation suggests the use of \textbf{Monte Carlo methods} to solve for unknown function $h$. Monte Carlo methods are a class of numerical techniques based on simulating random variables used to solve a range of problems, such as numerical integration and optimization. \\

Returning to the theorem, let us now discuss its statement:
\begin{itemize}
	\item When confronted with a PDE of the form above, we can define a (fictitious) process $X$ with drift and volatility given by the processes $a(t,X_t)$ and $b(t,X_t)$, respectively.
	\item Thinking of $c$ as a ``discount factor,'' we then consider the conditional expectation of the discounted terminal condition $H(X_T)$ and the running term $g(t,X_t)$ given that the value of $X$ at time $t$ is equal to a known value, $x$. Clearly, this conditional expectation is a function of $t$ and $x$; for every value of $t$ and $x$ we have some conditional expectation value.
	\item This function (the conditional expectation as a function of $t$ and $x$) is precisely the solution to the PDE we started with and can be estimated via Monte Carlo simulation of the process $X$. 
\end{itemize}

A class of Monte Carlo methods have also been developed for nonlinear PDEs, but this is beyond the scope of this work.

%% file: MachineLearning-NeuralNetworks.tex
\chapter{An Introduction to Deep Learning}
\setcounter{section}{4}


The tremendous strides made in computing power and the explosive growth in data collection and availability in recent decades has coincided with an increased interest in the field of \textbf{machine learning} (ML). This has been reinforced by the success of machine learning in a wide range of applications ranging from image and speech recognition, medical diagnostics, email filtering, fraud detection and many more. \\ 

\begin{figure}[h!]
	\centering
	\includegraphics[width=0.8\textwidth,trim={0 1.5em 0 1em},clip]{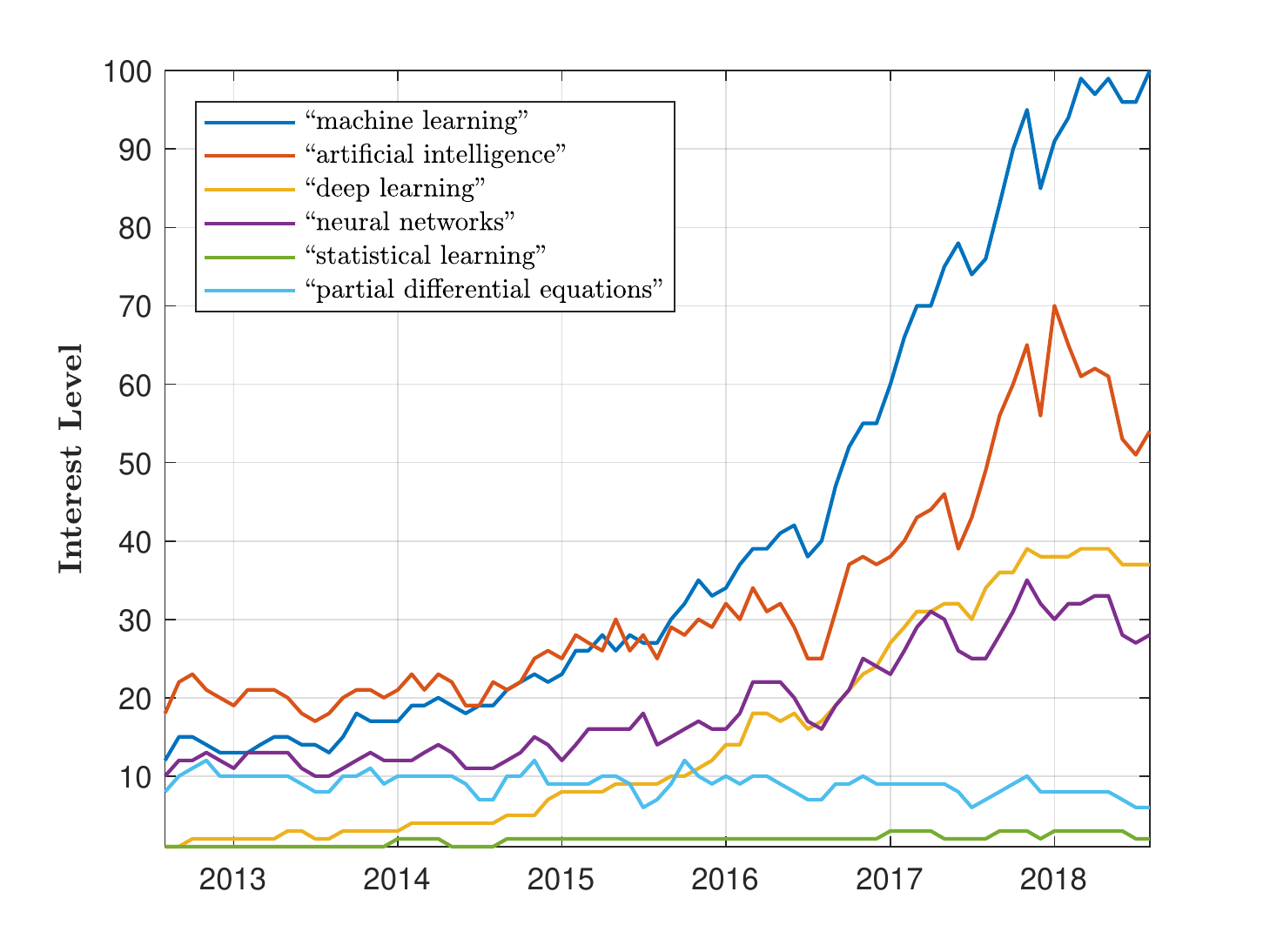}
	\caption{Google search frequency for various terms. A value of 100 is the peak popularity for the term; a value of 50 means that the term is half as popular. }
	\label{fig:ML_pop}
\end{figure}

As the name suggests, the term machine learning refers to computer algorithms that learn from data. The term ``learn'' can have several meanings depending on the context, but the the common theme is the following: a computer is faced with a task and an associated performance measure, and its goal is to improve its performance in this task with experience which comes in the form of examples and data. \\

ML naturally divides into two main branches. \textbf{Supervised learning} refers to the case where the data points include a label or target and tasks involve predicting these  labels/targets (i.e. classification and regression). \textbf{Unsupervised learning} refers to the case where the dataset does not include such labels and the task involves learning a useful structure that relates the various variables of the input data (e.g. clustering, density estimation). Other branches of ML, including semi-supervised and reinforcement learning, also receive a great deal of research attention at present. For further details the reader is referred to \cite{bishop2006pattern} or \cite{goodfellow2016deep}. \\

An important concept in machine learning is that of \textbf{generalization} which is related to the notions of \textbf{underfitting} and \textbf{overfitting}. In many ML applications, the goal is to be able to make meaningful statements concerning data that the algorithm has not encountered - that is, to generalize the model to unseen examples. It is possible to calibrate an assumed model ``too well'' to the training data in the sense that the model gives misguided predictions for new data points; this is known as overfitting. The opposite case is underfitting, where the model is not fit sufficiently well on the input data and consequently does not generalize to test data. Striking a balance in the trade-off between underfitting and overfitting, which itself can be viewed as a tradeoff between bias and variance, is crucial to the success of a ML algorithm. \\
	
On the theoretical side, there are a number of interesting results related to ML. For example, for certain tasks and hypothesized models it may be possible to obtain the minimal sample size to ensure that the training error is a faithful representation of the generalization error with high confidence (this is known as \textbf{Probably Approximately Correct (PAC) learnability}). Another result is the \textbf{no-free-lunch theorem}, which implies that there is no universal learner, i.e. that every learner has a task on which it fails even though another algorithm can successfully learn the same task. For an excellent exposition of the theoretical aspects of machine learning the reader is referred to \cite{shalev2014understanding}.

\subsection{Neural Networks and Deep Learning}
\textbf{Neural networks} are machine learning models that have received a great deal of attention in recent years due to their success in a number of different applications. The typical way of motivating the approach behind neural network models is to compare the way they operate to the human brain. The building blocks of the brain (and neural networks) are basic computing devices called \textbf{neurons} that are connected to one another by a complex communication network. The communication links cause the activation of a neuron to activate other neurons it is connected to. From the perspective of learning, training a neural network can be thought of as determining which neurons ``fire'' together. \\

Mathematically, a neural network can be defined as a directed graph with vertices representing neurons and edges representing links. The input to each neuron is a function of a weighted sum of the output of all neurons that are connected to its incoming edges. There are many variants of neural networks which differ in architecture (how the neurons are connected); see \hyperref[fig:NN_arch]{Figure \ref{fig:NN_arch}}. The simplest of these forms is the \textbf{feedforward neural network}, which is also referred to as a \textbf{multilayer perceptron} (MLP). \\

\afterpage{
	\begin{figure}[h!]
		\centering
		\includegraphics[width=0.85\textwidth,trim={0 3em 0 3em},clip]{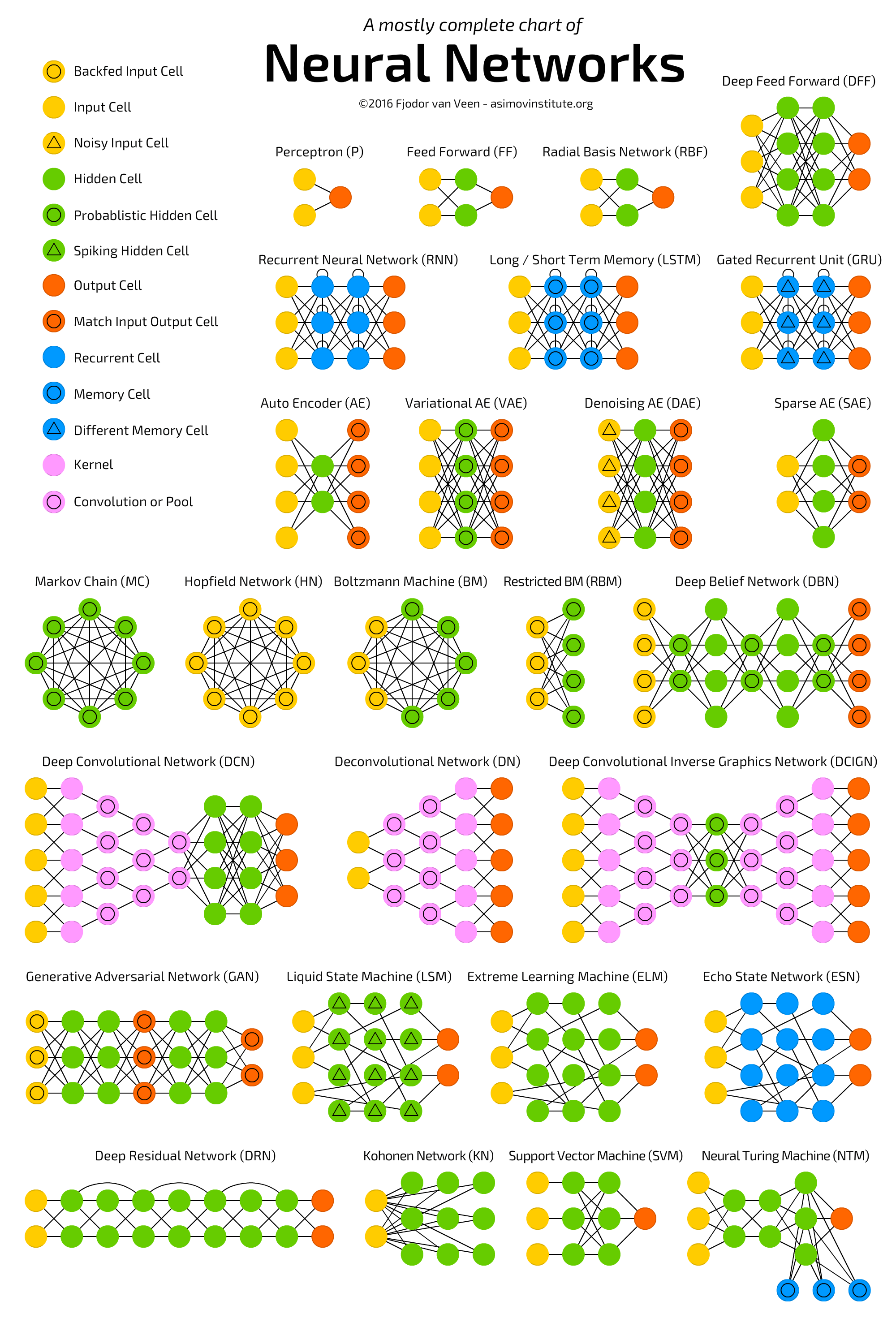}
		\caption{Neural network architectures. Source: ``Neural Networks 101'' by Paul van der Laken (\url{https://paulvanderlaken.com/2017/10/16/neural-networks-101})}
		\label{fig:NN_arch}
	\end{figure}
	\newpage
}

MLPs can be represented by a directed acyclic graph and as such can be seen as feeding information forward. Usually, networks of this sort are described in terms of layers which are chained together to create the output function, where a layer is a collection of neurons that can be thought of as a unit of computation. In the simplest case, there is a single \textbf{input layer} and a single \textbf{output layer}. In this case, output $j$ (represented by the $j$th neuron in the output layer), is connected to the input vector $\bx$ via a biased weighted sum and an \textbf{activation function} $\phi_j$:
\[ y_j = \phi_j \left( b_j + \sum_{i=1}^d w_{i,j} \bx_i  \right) \]

It is also possible to incorporate additional \textbf{hidden layers} between the input and output layers. For example, with one hidden layer the output would become: 
\[y_k = \redunderbrace{\phi \left[ b_k^{(2)} + \sum_{i=1}^{m_2} \ w_{j,k}^{(2)} \cdot \redunderbrace{\psi \left(b_j^{(1)} + \sum_{i=1}^{m_1} w_{i,j}^{(1)} \bx_j \right) }{\mbox{input layer to hidden layer}}  \right]}{hidden layer to output layer} \]
where $\phi,\psi: \mathbb{R} \to \mathbb{R}$ are nonlinear activation functions for each layer and the bracketed superscripts refer to the layer in question. We can visualize an extension of this the process as a simple application of the chain rule, e.g.
\[f(\bx) = \psi_d( \cdots \psi_2(\psi_1(\bx)))\]
Here, each layer of the network is represented by a function $\psi_i$, incorporating the weighted sums of previous inputs and activations to connected outputs. The number of layers in the graph is referred to as the \textbf{depth} of the neural network and the number of neurons in a layer represents the \textbf{width} of that particular layer; see \hyperref[fig:MLP]{Figure \ref{fig:MLP}}. \\

The term ``deep'' neural network and \textbf{deep learning} refer to the use of neural networks with many hidden layers in ML problems. One of the advantages of adding hidden layers is that depth can exponentially reduce the computational cost in some applications and exponentially decrease the amount of training data needed to learn some functions. This is due to the fact that some functions can be represented by smaller deep networks compared to wide shallow networks. This decrease in model size leads to improved statistical efficiency. \\

\def\layersep{2.5cm}

\begin{figure}
	\centering
\begin{tikzpicture}[shorten >=1pt,->,draw=black!50, node distance=\layersep]
\tikzstyle{every pin edge}=[<-,shorten <=1pt]
\tikzstyle{neuron}=[circle,fill=black!25,minimum size=17pt,inner sep=0pt]
\tikzstyle{input neuron}=[neuron, fill=green!50];
\tikzstyle{output neuron}=[neuron, fill=red!50];
\tikzstyle{hidden neuron}=[neuron, fill=blue!50];
\tikzstyle{annot} = [text width=4em, text centered]

\foreach \name / \y in {1,...,4}
\node[input neuron, pin=left:Input \y] (I-\name) at (0,-\y) {};

\foreach \name / \y in {1,...,5}
\path[yshift=0.5cm]
node[hidden neuron] (H-\name) at (\layersep,-\y cm) {};

\node[output neuron,pin={[pin edge={->}]right:Output}, right of=H-3] (O) {};

\foreach \source in {1,...,4}
\foreach \dest in {1,...,5}
\path (I-\source) edge (H-\dest);

\foreach \source in {1,...,5}
\path (H-\source) edge (O);

\node[annot,above of=H-1, node distance=1cm] (hl) {Hidden layer};
\node[annot,left of=hl] {Input layer};
\node[annot,right of=hl] {Output layer};
\end{tikzpicture}
\caption{Feedforward neural network with one hidden layer.}
\label{fig:MLP}
\end{figure}
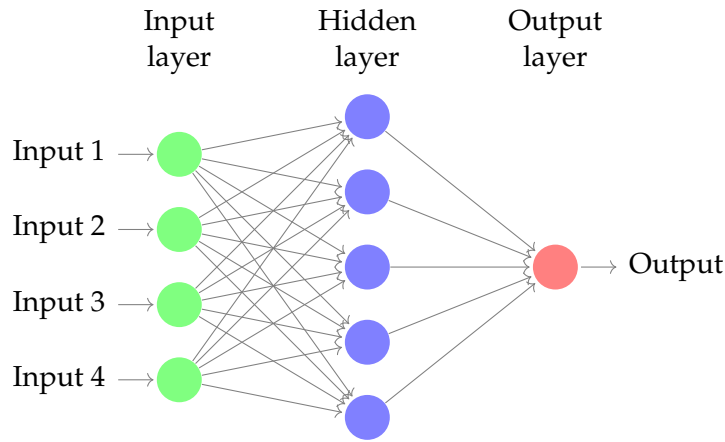


It is easy to imagine the tremendous amount of flexibility and complexity that can be achieved by varying the structure of the neural network. One can vary the depth or width of the network, or have varying activation functions for each layer or even each neuron. This flexibility can be used to achieve very strong results but can lead to opacity that prevents us from understanding why any strong results are being achieved. \\

Next, we turn to the question of how the parameters of the neural network are estimated. To this end, we must first define a \textbf{loss function}, $L(\vtheta;\bx,\by)$, which will determine the performance of a given parameter set $\vtheta$ for the neural network consisting of the weights and bias terms in each layer. The goal is to find the parameter set that minimizes our loss function. The challenge is that the highly nonlinear nature of neural networks can lead to non-convexities in the loss function. Non-convex optimization problems are non-trivial and often we cannot guarantee that a candidate solution is a global optimizer.

\subsection{Stochastic Gradient Descent}

The most commonly used approach for estimating the parameters of a neural network is based on \textbf{gradient descent} which is a simple methodology for optimizing a function. Given a function $f: \RR^d \rightarrow \RR$, we wish to determine the value of $\vx$ that achieves the minimum value of $f$. To do this, we begin with an initial guess $\vx_0$ and compute the gradient of $f$ at this point. This gives the direction in which the largest increase in the function occurs. To minimize the function we move in the opposite direction, i.e. we iterate according to:
\[ \vx_{n} = \vx_{n-1} - \eta \cdot \nabla_\vx f\left(\vx_{n-1}\right)   \]
where $\eta$ is the step size known as the \textbf{learning rate}, which can be constant or decaying in $n$. The algorithm converges to a critical point when the gradient is equal to zero, though it should be noted that this is not necessarily a global minimum. In the context of neural networks, we would compute the derivatives of the loss functional with respect to the parameter set $\vtheta$ (more on this in the next section) and follow the procedure outlined above. \\

One difficulty with the use of gradient descent to train neural networks is the computational cost associated with the procedure when training sets are large. This necessitates the use of an extension of this algorithm known as \textbf{stochastic gradient descent} (SGD). When the loss function we are minimizing is additive, it can be written as:
\[ \nabla L(\vtheta;\bx,\by) = \frac{1}{m} \sum_{i=1}^m \nabla_\vtheta L_i \left( \vtheta; \bx^{(i)},\by^{(i)} \right)  \]   
where $m$ is the size of the training set and $L_i$ is the per-example loss function. The approach in SGD is to view the gradient as an expectation and approximate it with a random subset of the training set called a \textbf{mini-batch}. That is, for a fixed mini-batch of size $m'$ the gradient is estimated as: 
\[ \nabla_\vtheta L(\vtheta;\bx,\by) \approx \frac{1}{m'} ~ \nabla_\vtheta \sum_{i=1}^{m'} L_i \left( \vtheta; \bx^{(i)},\by^{(i)} \right) \]
This is followed by taking the usual step in the opposite direction (steepest descent). \\

\subsection{Backpropagation}

The stochastic gradient descent optimization approach described in the previous section requires repeated computation of the gradients of a highly nonlinear function. \textbf{Backpropagation} provides a computationally efficient means by which this can be achieved. It is based on recursively applying the chain rule and on defining computational graphs to understand which computations can be run in parallel. \\

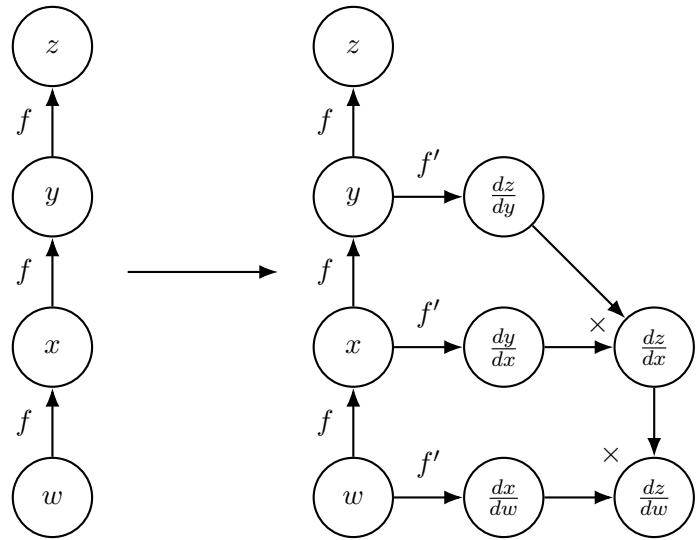
\begin{figure}[h!]
	\centering
	\begin{tikzpicture}
	
	\def\layersepx{2.0cm}
	\def\layersepy{2.0cm}
	\tikzstyle{neuron}=[circle, draw=black, minimum size=30pt, thick]
	\tikzstyle{arrow}=[draw=black, -{Latex[length=2.5mm]}, thick]
	\tikzstyle{fleft}=[left,inner sep=7pt]
	\tikzstyle{ftop}=[above,inner sep=7pt]
	
	\node[neuron] (n00) at (0*\layersepx,0*\layersepy) {$w$};
	\node[neuron] (n01) at (0*\layersepx,1*\layersepy) {$x$};
	\node[neuron] (n02) at (0*\layersepx,2*\layersepy) {$y$};
	\node[neuron] (n03) at (0*\layersepx,3*\layersepy) {$z$};
	
	\draw[arrow] (n00) -- node[fleft]{$f$} (n01);
	\draw[arrow] (n01) -- node[fleft]{$f$} (n02);
	\draw[arrow] (n02) -- node[fleft]{$f$} (n03);

	\node[neuron] (n20) at (2*\layersepx,0*\layersepy) {$w$};
	\node[neuron] (n21) at (2*\layersepx,1*\layersepy) {$x$};
	\node[neuron] (n22) at (2*\layersepx,2*\layersepy) {$y$};
	\node[neuron] (n23) at (2*\layersepx,3*\layersepy) {$z$};
	
	\draw[arrow] (n20) -- node[fleft]{$f$} (n21);
	\draw[arrow] (n21) -- node[fleft]{$f$} (n22);
	\draw[arrow] (n22) -- node[fleft]{$f$} (n23);

	\node[neuron] (n30) at (3*\layersepx,0*\layersepy) {$\frac{dx}{dw}$};
	\node[neuron] (n31) at (3*\layersepx,1*\layersepy) {$\frac{dy}{dx}$};
	\node[neuron] (n32) at (3*\layersepx,2*\layersepy) {$\frac{dz}{dy}$};
	
	\draw[arrow] (n20) -- node[ftop]{$f'$} (n30);
	\draw[arrow] (n21) -- node[ftop]{$f'$} (n31);
	\draw[arrow] (n22) -- node[ftop]{$f'$} (n32);

	\node[neuron] (n40) at (4*\layersepx,0*\layersepy) {$\frac{dz}{dw}$};
	\node[shift=(135:23pt)] at (4*\layersepx,0*\layersepy) {$\times$};
	\node[neuron] (n41) at (4*\layersepx,1*\layersepy) {$\frac{dz}{dx}$};
	\node[shift=(158:23pt)] at (4*\layersepx,1*\layersepy) {$\times$};
	
	\draw[arrow] (n30) -- (n40);
	\draw[arrow] (n31) -- (n41);
	\draw[arrow] (n32) -- (n41);
	\draw[arrow] (n41) -- (n40);

	\draw[arrow] (0.5*\layersepx,1.5*\layersepy) -- (1.5*\layersepx,1.5*\layersepy);

	\end{tikzpicture}
	\caption{Visualization of backpropagation algorithm via computational graphs. The left panel shows the composition of functions connecting input to output; the right panel shows the use of the chain rule to compute the derivative. Source: \cite{goodfellow2016deep}}
	\label{fig:backprop_figure_2}
\end{figure}

As we have seen in previous sections, a feedforward neural network can be thought of as receiving an input $\mathbf{x}$ and computing an output $y$ by evaluating a function defined by a sequence of compositions of simple functions. These simple functions can be viewed as operations between nodes in the neural network graph. With this in mind, 
the derivative of $y$ with respect to $\mathbf{x}$ can be computed analytically by 
repeated applications of the chain rule, given enough information about the 
operations between nodes. The backpropagation algorithm traverses the graph, repeatedly computing the chain rule until the derivative of the output $y$ with respect to $\mathbf{x}$ is represented symbolically via a second computational graph; see \hyperref[fig:backprop_figure_2]{Figure \ref{fig:backprop_figure_2}}.  \\

\vspace{0.5cm}

\vspace{0.5cm}


The two main approaches for computing the derivatives in the computational graph is to input a numerical value then compute the derivatives at this value, returning a number as done in PyTorch (\url{pytorch.org}), or to compute the derivatives of a symbolic variable, then store the derivative operations into new nodes added to the graph for later use as done in TensorFlow (\url{tensorflow.org}). The advantage of the latter approach is that higher-order derivatives can be computed from this extended graph by running backpropagation again. \\

The backpropagation algorithm takes at most $\mathcal{O}\left(n^2\right)$ operations 
for a graph with $n$ nodes, storing at most $\mathcal{O}\left(n^2\right)$ new nodes. In practice, 
most feedforward neural networks are designed in a chain-like way, which in turn reduces  
the number of operations and new storages to $\mathcal{O}\left(n\right)$, making derivatives calculations 
a relatively cheap operation. \\

\subsection{Summary}

In summary, training neural networks is broadly composed of three ingredients: 
\begin{enumerate}
	\item Defining the architecture of the neural network and a loss function, also known as the \textbf{hyperparameters} of the model;
	\item Finding the loss minimizer using stochastic gradient descent;
	\item Using backpropagation to compute the derivatives of the loss function. \\
\end{enumerate}  

This is presented in more mathematical detail in \hyperref[fig:algo]{Figure \ref{fig:algo}}. \\

\begin{figure}[h!]
	{\small
		\begin{tabular}{r}
			\hline\hline
			\hspace{0.95\textwidth}
		\end{tabular}
		\vspace{-1em}
		\begin{enumerate}
			\item Define the architecture of the neural network by setting its depth (number \\
			of layers), width (number of neurons in each layer) and activation functions
				
			\item Define a loss functional $L(\vtheta; \bx, \by)$, mini-batch size $m'$ and learning rate $\eta$
			
			\item Minimize the loss functional to determine the optimal $\vtheta$:
			\begin{enumerate} 
				
				\item \texttt{Initialize the parameter set, $\vtheta_0$}
				\item \texttt{Randomly sample a mini-batch of $m'$ training examples $\left( \bx^{(i)}, \by^{(i)} \right)$}
				\item \texttt{Compute the loss functional for the sampled mini-batch $L\left(\vtheta_i; \bx^{(i)}, \by^{(i)} \right)$}
				\item \texttt{Compute the gradient $\nabla_\vtheta L \left( \vtheta_i; \bx^{(i)}, \by^{(i)} \right)$ using backpropagation }
				\item \texttt{Use the estimated gradient to update $\theta_i$ based on SGD:
				\[ \vtheta_{i+1} = \vtheta_i - \eta \cdot \nabla_\vtheta L(\vtheta_i; \bx^{(i)}, \by^{(i)}) \] }
				\item \texttt{Repeat steps (b)-(e) until $\|\vtheta_{i+1} - \vtheta_i\|$ is small. }
			\end{enumerate}
		\end{enumerate}
		\begin{tabular}{r}
			\hline\hline
			\hspace{0.95\textwidth}
		\end{tabular}
	}
	\vspace{-1em}
	\caption{Parameter estimation procedure for neural networks.} \label{fig:algo}
\end{figure}

\subsection{The Universal Approximation Theorem}

An important theoretical result that sheds some light on why neural networks perform well is the \textbf{universal approximation theorem}, see \cite{cybenko1989approximation} and \cite{hornik1991approximation}. In simple terms, this result states that any continuous function defined on a compact subset of $\RR^n$ can be approximated arbitrarily well by a feedforward network with a single hidden layer. \\

Mathematically, the statement of the theorem is as follows: let $\phi$ be a nonconstant, bounded, monotonically-increasing continuous function and let $I_m$ denote the $m$-dimensional unit hypercube. Then, given any $\epsilon > 0$ and any function $f$ defined on $I_m$, there exists $N, v_i, b_i, \vw$ such that the approximation function:
\[ F(\vx) = \sum_{i=1}^{N} v_i \phi\left( \vw \cdot \vx + \vb_i \right) \]
satisfies $|F(\vx) - f(\vx)| < \epsilon$ for all $\vx \in I_m$. \\

A remarkable aspect of this result is the fact that the activation function is independent of the function we wish to approximate! However, it should be noted that the theorem makes no statement on the number of neurons needed in the hidden layer to achieve the desired approximation error, nor whether the estimation of the parameters of this network is even feasible. \\

\subsection{Other Topics}

\subsubsection{Adaptive Momentum}
Recall that the stochastic gradient descent algorithm is parametrized by a learning rate $\eta$ which determines the step size in the direction of steepest descent given by the gradient vector. In practice, this value should decrease along successive iterations of the SGD algorithm for the network to be properly trained. For a network's parameter set to be properly optimized, an appropriately chosen learning rate schedule is in order, as it ensures that the excess error is decreasing in each iteration. Furthermore, this learning rate schedule can depend on the nature of the problem at hand. \\

For the reasons discussed in the last paragraph, a number of different algorithms have been developed to find some heuristic capable of guiding the selection of an effective sequence of learning rate parameters. Inspired by physics, many of these algorithms interpret the gradient as a velocity vector, that is, the direction and speed at which the parameters move through the parameter space. \textbf{Momentum algorithms}, for example, calculate the next velocity as a weighted sum of the gradient from the last iteration and the newly calculated one. This helps minimize instabilities caused by the high sensitivity of the loss function with respect to some directions of the parameter space, at the cost of introducing two new parameters, namely a decay \textbf{factor}, and an initialization parameter $\eta_0$. Assuming these sensitivities are axis-dependent, we can apply different learning rate schedules to each direction and adapt them throughout the training session. \\

The work of \cite{kingma2014adam} combines the ideas discussed in this section in a single framework referred to as \textbf{Adaptative Momentum} (Adam). The main idea is to increase/decrease the learning rates based on the past magnitudes of the partial derivatives for a particular direction. Adam is regarded as being robust to its hyperparameter values. \\

\subsubsection{The Vanishing Gradient Problem}

In our analysis of neural networks, we have established that the addition of layers to a network's architecture can potentially lead to great increases in its performance: increasing the number of layers allows the network to better approximate increasingly more complicated functions in a more efficient manner. In a sense, the success of deep learning in current ML applications can be attributed to this notion. \\

However, this improvement in power can be counterbalanced by the \textbf{Vanishing Gradient Problem}: due to the the way gradients are calculated by backpropagation, the deeper a network is the smaller its loss function's derivative with respect to weights in early layers becomes. At the limit, depending on the activation function, the gradient can underflow in a manner that causes weights to not update correctly. \\ 

Intuitively, imagine we have a deep feedforward neural network consisting of $n$ layers. At every iteration, each of the network's weights receives an update that is proportional to the gradient of the error function with respect to the current weights. As these gradients are calculated using the chain rule through backpropagation, the further back a layer is, the more it is multiplied by an already small gradient. \\

\subsubsection{Long-Short Term Memory and Recurrent Neural Networks}

Applications with time or positioning dependencies, such as speech recognition and natural language processing, where each layer of the network handles one time/positional step, are particularly prone to the vanishing gradient problem. In particular, the vanishing gradient might mask long term dependencies between observation points far apart in time/space.  \\

Colloquially, we could say that the neural network is not able to accurately remember important information from past layers. One way of overcoming this difficulty is to incorporate a notion of memory for the network, training it to learn which inputs from past layers should flow through the current layer and pass on to the next, i.e. how much information should be ``remembered'' or ``forgotten.'' This is the intuition behind \textbf{long-short term memory} (LSTM) networks, introduced by \cite{hochreiter1997long}. \\

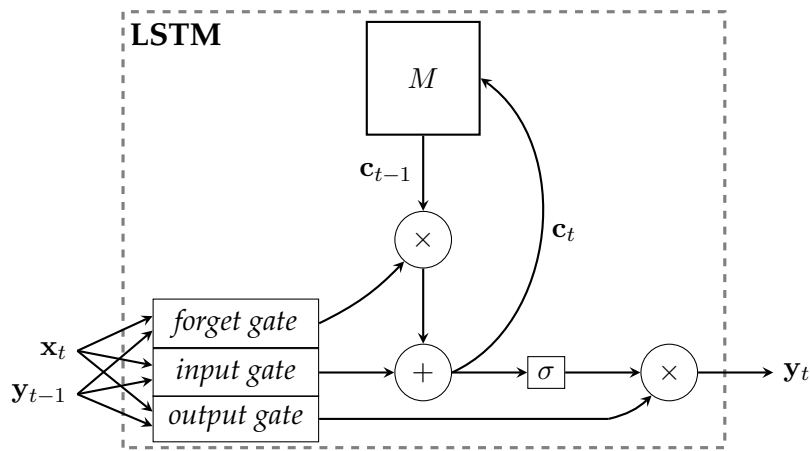
\begin{figure}[h!]
\centering

\begin{tikzpicture}

\draw[-stealth, very thick, dashed, gray] (-1.5, -1.0) rectangle (6.5, 4.8);
\node[] (tttxt) at (-0.8, 4.5) {\large \bf LSTM};

\node[rectangle, draw, minimum width=2.2cm] (IG) {\emph{input gate}};
\node[rectangle, above=+0.00em of IG, draw, minimum width=2.2cm] (FG) {\emph{forget gate}};
\node[rectangle, below=-0.05em of IG, draw, minimum width=2.2cm] (OG) {\emph{output gate}};
\node[left=of IG, yshift=0.75em)] (X) {$\vec{x}_t$};
\node[left=of IG, yshift=-0.75em)] (Y) {$\vec{y}_{t-1}$};

\draw[-stealth, thick] (X.east) -- ([yshift=0.25em]IG.west);
\draw[-stealth, thick] (X.east) -- ([yshift=0.25em]FG.west);
\draw[-stealth, thick] (X.east) -- ([yshift=0.25em]OG.west);
\draw[-stealth, thick] (Y.east) -- ([yshift=-0.25em]IG.west);
\draw[-stealth, thick] (Y.east) -- ([yshift=-0.25em]FG.west);
\draw[-stealth, thick] (Y.east) -- ([yshift=-0.25em]OG.west);

\node[circle, draw, right=of IG] (pl) {$+$};
\node[rectangle, draw, right=of pl] (th) {$\sigma$};
\node[circle, draw, right=of th] (t2) {$\times$};
\node[right=of t2] (Y1) {$\vec{y}_t$};

\node[circle, draw, above=of pl] (t3) {$\times$};

\node[rectangle, thick, draw, above=of t3, minimum width=1.5cm, minimum height=1.5cm] (M) {$M$};

\draw[-stealth, thick] (IG) -- (pl);
\draw[-stealth, thick] (pl) -- (th);
\draw[-stealth, thick] (th) -- (t2);
\draw[-stealth, thick] (t2) -- (Y1);
\draw[-stealth, thick] (M) -- node[left] {$\vec{c}_{t-1}$} (t3);
\draw[-stealth, thick] (t3) -- (pl);
\draw[thick] (OG.east) -- ([xshift=10em]OG.east);
\path[-stealth, thick] (OG.east) -- ([xshift=10em]OG.east) edge[bend right=15] (t2);
\path[-stealth, thick] (FG.east) edge[bend right=10] (t3);
\path[-stealth, thick] (pl.east) edge[bend right=60] node[right] {$\vec{c}_t$} (M.east);

\end{tikzpicture}
\caption{Architecture of an LSTM unit: a new input $\vec{x}_t$ and output of the last unit $\vec{y}_{t-1}$ are combined with past memory information $\vec{c}_{t-1}$ to produce new output $\vec{y}_{t}$ and store new memory information $\vec{c}_{t}$.  Source: ``A trip down long-short memory lane'' by Peter Veli\v{c}kovi\'{c} (\url{https://www.cl.cam.ac.uk/~pv273/slides/LSTMslides.pdf})}
\label{fig:LSTM}
\end{figure}

LSTM networks are a class of \textbf{recurrent neural networks} (RNNs) that consists of layers called \textbf{LSTM units}. Each layer is composed of a \textbf{memory cell}, an \textbf{input gate}, an \textbf{output gate} and a \textbf{forget gate} which regulates the flow of information from one layer to another and allows the network to learn the optimal remembering/forgetting mechanism. Mathematically, some fraction of the gradients from past layers are able to pass through the current layer directly to the next. The magnitude of the gradient that passes through the layer unchanged (relative to the portion that is transformed) as well as the discarded portion, is also learned by the network. This embeds the memory aspect in the  architecture of the LSTM allowing it to circumvent the vanishing gradient problem and learn long-term dependencies; refer to \hyperref[fig:LSTM]{Figure \ref{fig:LSTM}} for a visual representation of a single LSTM unit. \\

\vspace{0.5cm}

\vspace{0.5cm}

Inspired by the effectiveness of LSTM networks and given the rising importance of deep architectures in modern ML, \cite{srivastava2015highway} devised a network that allows gradients from past layers to flow through the current layer. \textbf{Highway networks} use the architecture of LSTMs for problems where the data is not sequential. By adding an ``information highway,'' which allows gradients from early layers to flow unscathed through intermediate layers to the end of the network, the authors are able to train incredibly deep networks, with depth as high as a 100 layers without vanishing gradient issues. \\

%% file: DGM.tex
\chapter{The Deep Galerkin Method} \label{chap:DGM}
\setcounter{section}{5}

\subsection{Introduction}
We now turn our attention to the application of neural networks to finding solutions to PDEs.
As discussed in \hyperref[chap:numerical]{Chapter \ref{chap:numerical}}, numerical methods that are based on grids can fail when the dimensionality of the problem becomes too large. In fact, the number of points in the mesh grows exponentially in the number of dimensions which can lead to computational intractability. Furthermore, even if we were to assume that the computational cost was manageable, ensuring that the grid is set up in a way to ensure stability of the finite difference approach can be cumbersome. \\

With this motivation, \cite{sirignano2017dgm} propose a \textit{mesh-free} method for solving PDEs using neural networks. The \textbf{Deep Galerkin Method} (DGM) approximates the solution to the PDE of interest with a deep neural network. With this parameterization, a loss function is set up to penalize the fitted function's deviations from the desired differential operator and boundary conditions.  The approach takes advantage of computational graphs and the backpropagation algorithm discussed in the previous chapter to efficiently compute the differential operator while the boundary conditions are straightforward to evaluate. For the training data, the network uses points randomly sampled from the region where the function is defined and the optimization is performed using stochastic gradient descent. \\

The main insight of this approach lies in the fact that the training data consists of randomly sampled points in the function's domain. By sampling mini-batches from different parts of the domain and processing these small batches sequentially, the neural network ``learns'' the function without the computational bottleneck present with grid-based methods. This circumvents the curse of dimensionality which is encountered with the latter approach.

\subsection{Mathematical Details}

The form of the PDEs of interest are generally described as follows: let $u$ be an unknown function of time and space defined on the region $[0,T] \times \Omega$ where $\Omega \subset \RR^d$, and assume that $u$ satisfies the PDE:

\begin{equation*}
\begin{cases}
\left(\partial_t + \LL\right) u(t,\vx) = 0, & \qquad (t,\vx) \in [0,T] \times \Omega 
\\
u(0,\vx) = u_0(\vx), & \qquad   \vx \in \Omega \hspace{2.47cm} \qquad  \text{\color{red} (initial condition)}
\\
u(t,\vx) = g(t,\vx), & \qquad  (t,\vx) \in [0,T] \times \partial\Omega \hspace{0.3cm} \qquad \text{\color{red} (boundary condition)}
\end{cases}
\end{equation*} \\

The goal is to approximate $u$ with an approximating function $f(t,\vx;\vtheta)$ given by a deep neural network with parameter set $\vtheta$. The loss functional for the associated training problem consists of three parts:

\begin{enumerate}
	\item A measure of how well the approximation satisfies the \textbf{differential operator}: 
	\[ \Big\| \left(\partial_t + \LL\right) f(t,\vx; \vtheta) \Big\|_{[0,T] \times \Omega, ~\nu_1}^2 \]
	\textit{Note: parameterizing $f$ as a neural network means that the differential operator can be computed easily using backpropagation.}
	\item A measure of how well the approximation satisfies the \textbf{boundary condition}:
	\[ \Big\|f(t,\vx; \vtheta) - g(t,\vx)\Big\|_{[0,T] \times \partial\Omega, ~\nu_2}^2 \]
	\item A measure of how well the approximation satisfies the \textbf{initial condition}:
	\[ \Big\|f(0,\vx; \vtheta) - u_0(\vx)\Big\|_{\Omega, ~\nu_3}^2 \qquad ~ \]
\end{enumerate}

\vspace{0.3cm}

In all three terms above the error is measured in terms of $L^2$-norm, i.e. using $\big\| h(y) \big\|_{\mathcal{Y}, \nu}^2 = \int_{\mathcal{Y}}|h(y)|^2\nu(y)dy$ with $\nu(y)$ being a density defined on the region $\mathcal{Y}$. Combining the three terms above gives us the cost functional associated with training the neural network:

{\footnotesize \begin{equation*} 
L(\vtheta) = \redunderbrace{\Big\| \left(\partial_t + \LL\right) f(t,\vx; \vtheta) \Big\|_{[0,T] \times \Omega, \nu_1}^2}{\mbox{differential operator}} 
+ 
\redunderbrace{\Big\|f(t,\vx; \vtheta) - g(t,\vx)\Big\|_{[0,T] \times \partial\Omega, \nu_2}^2}{\mbox{boundary condition}}
+
\redunderbrace{\Big\|f(0,\vx; \vtheta) - u_0(\vx)\Big\|_{\Omega, \nu_3}^2}{\mbox{initial condition}}
\end{equation*} 
}

The next step is to minimize the loss functional using stochastic gradient descent. More specifically, we apply the algorithm defined in Figure \ref{fig:DGM}. The description given in Figure \ref{fig:DGM} should be thought of as a general outline as the algorithm should be modified according to the particular nature of the PDE being considered. \\

\begin{figure}[H]
	{\small
		\begin{tabular}{r}
			\hline\hline
			\hspace{0.975\textwidth}
		\end{tabular}
		\vspace{-1.5em}
		\begin{enumerate}
			\item Initialize the parameter set $\vtheta_0$ and the learning rate $\alpha_n$. 
			\item Generate random samples from the domain's interior and time/spatial boundaries, i.e.
			\begin{itemize}
				\item \texttt{Generate $(t_n,x_n)$ from $[0,T] \times \Omega $ according to $\nu_1$} 
				\item \texttt{Generate $(\tau_n, z_n)$ from $[0,T] \times \partial\Omega$ according to $\nu_2$}
				\item \texttt{Generate $w_n$ from $\Omega$, according to $\nu_3$}
			\end{itemize} 
			\item Calculate the loss functional for the current mini-batch (the randomly sampled points $s_n = \left\{(t_n, x_n), (\tau_n, z_n), w_n\right\}$): 
			\begin{itemize}
				\item \texttt{Compute $L_1(\vtheta_n;t_n,x_n) = \left(\left(\partial_t + \LL\right) f(\vtheta_n;t_n,x_n) \right)^2 $ } 
				\item \texttt{Compute $L_2(\vtheta_n;\tau_n,z_n) = (f(\tau_n,z_n) - g(\tau_n,z_n))^2 $ } 
				\item \texttt{Compute $L_3(\vtheta_n;w_n) = (f(0,w_n) - u_0(w_n))^2 $ } 
				\item \texttt{Compute $L(\vtheta_n;s_n) = L_1(\vtheta_n;t_n,x_n) + L_2(\vtheta_n;\tau_n,z_n) + L_3(\vtheta_n;z_n) $ } 
			\end{itemize} 
			\item Take a descent step at the random point $s_n$ with Adam-based learning rates:
			\[\theta_{n+1} = \theta_n - \alpha_n \nabla_\vtheta L(\vtheta_n; s_n)\]
			\item Repeat steps (2)-(4) until $\|\theta_{n+1} - \theta_n \|$ is small.
		\end{enumerate}
		\begin{tabular}{r}
			\hline\hline
			\hspace{0.975\textwidth}
		\end{tabular}
	}
	\vspace{-1em}
	\caption{Deep Galerkin Method (DGM) algorithm. \label{fig:DGM}}
\end{figure}

\vspace{0.5cm}

It is important to notice that the problem described here is strictly an optimization problem. This is unlike typical machine learning applications where we are concerned with issues of underfitting, overfitting and generalization. Typically, arriving at a parameter set where the loss function is equal to zero would not be desirable as it suggests some form of overfitting. However, in this context a neural network that achieves this is the goal as it would be the solution to the PDE at hand. The only case where generalization becomes relevant is when we are unable to sample points everywhere within the region where the function is defined, e.g. for functions defined on unbounded domains. In this case, we would be interested in examining how well the function satisfies the PDE in those unsampled regions. The results in the next chapter suggest that this generalization is often very poor. \\

\subsection{A Neural Network Approximation Theorem}

Theoretical motivation for using neural networks to approximate solutions to PDEs is given as an elegant result in \cite{sirignano2017dgm} which is similar in spirit to the Universal Approximation Theorem. More specifically, it is shown that \textit{deep neural network approximators converge to the solution of a class of quasilinear parabolic PDEs as the number of hidden layers tends to infinity}. To state the result in more precise mathematical terms, define the following:
\begin{itemize}
	\item $L(\vtheta)$, the loss functional measuring the neural network's fit to the differential operator and boundary/initial/terminal conditions;
	\item $\mathfrak{C}^n$, the class of neural networks with $n$ hidden units;
	\item $f^n = \underset{f \in \mathfrak{C}^n}{\arg \min} ~ L(\vtheta)$, the best $n$-layer neural network approximation to the PDE solution.
\end{itemize} 
The main result is the convergence of the neural network approximators to the \textit{true} PDE solution:
\[ f^n \rightarrow u \qquad \text{as} \qquad n \rightarrow \infty \] 

\vspace{0.3cm}

Further details, conditions, statement of the theorem and proofs are found in Section 7 of \cite{sirignano2017dgm}. It is should be noted that, similar to the Universal Approximation Theorem, this result does not prescribe a way of designing or estimating the neural network successfully. \\

\subsection{Implementation Details}

The architecture adopted by \cite{sirignano2017dgm} is similar to that of LSTMs and Highway Networks described in the previous chapter. It consists of three layers, which we refer to as \textbf{DGM layers}: an input layer, a hidden layer and an output layer, though this can be easily extended to allow for additional hidden layers. \\

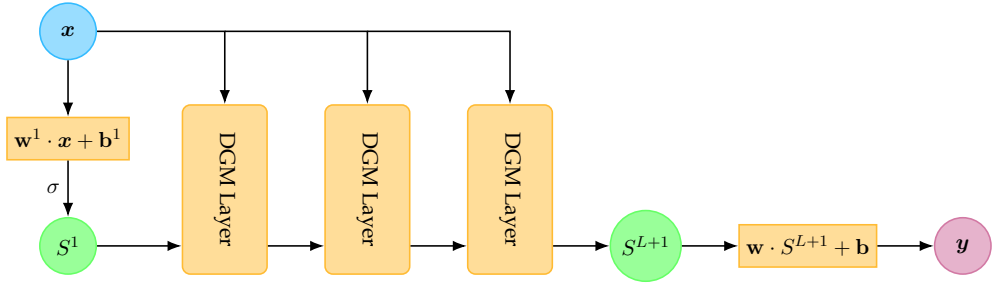
\begin{figure}[h!]
	\centering
	\scalebox{0.75}{
	\begin{tikzpicture}
	
	\definecolor{yellowd}{RGB}{255,187,51}
	\definecolor{yellowf}{RGB}{255,221,153}
	\definecolor{greend}{RGB}{102,255,102}
	\definecolor{greenf}{RGB}{153,255,153}
	\definecolor{blued}{RGB}{51,187,255}
	\definecolor{bluef}{RGB}{153,221,255}
	\definecolor{pinkd}{RGB}{210,121,164}
	\definecolor{pinkf}{RGB}{230,179,204}
	
	\tikzstyle{scircle}=[circle, thick, draw=greend, fill=greenf, minimum size=1cm]
	\tikzstyle{xcircle}=[circle, thick, draw=blued, fill=bluef, minimum size=1cm]
	\tikzstyle{ycircle}=[circle, thick, draw=pinkd, fill=pinkf, minimum size=1cm]
	\tikzstyle{smallrect}=[rectangle, thick, draw=yellowd, fill=yellowf, minimum height=0.75cm, minimum width=1.5cm,]
	\tikzstyle{dgmlayer}=[rectangle, thick, draw=yellowd, fill=yellowf, minimum height=3cm, minimum width=1.5cm, rounded corners]
	\tikzstyle{arrow}=[draw=black, -{Latex[length=2.5mm]}, thick]

	\node[smallrect] (R1) {$\vw^1 \cdot \vx + \vb^1$};
	\node[scircle, below=1.0cm of R1] (S1) {$S^1$};
	\node[xcircle, above=1.0cm of R1] (X) {$\vx$};
	
	\node[dgmlayer, right=1.5cm of S1, yshift=1cm] (L1) {};
	\node[rotate=-90] at (L1) {DGM Layer};
	\node[dgmlayer, right=1.0cm of L1] (L2) {};
	\node[rotate=-90] at (L2) {DGM Layer};
	\node[dgmlayer, right=1.0cm of L2] (L3) {};
	\node[rotate=-90] at (L3) {DGM Layer};
	
	\node[scircle, right=1.0cm of L3, yshift=-1cm)] (SL) {$S^{L+1}$};
	\node[smallrect, right=of SL] (RL) {$\vw \cdot S^{L+1} + \vb$};
	\node[ycircle, right=of RL] (Y) {$\vy$};
	
	\draw[arrow] (X) -- (R1);
	\draw[arrow] (R1) -- node[anchor=east] {$\sigma$} (S1);
	\draw[arrow] (S1) -- (S1-|L1.west);
	\draw[arrow] (S1-|L1.east) -- (S1-|L2.west);
	\draw[arrow] (S1-|L2.east) -- (S1-|L3.west);
	
	\draw[arrow] (X) -| coordinate (aL1) (L1);
	\draw[arrow] (aL1) -| coordinate (aL2) (L2);
	\draw[arrow] (aL2) -| (L3);
	
	\draw[arrow] (S1-|L3.east) -- (SL);
	\draw[arrow] (SL) -- (RL);
	\draw[arrow] (RL) -- (Y);

	\end{tikzpicture}
	}
	\caption{Bird's-eye perspective of overall DGM architecture.}
	\label{fig:networkoutside}
\end{figure}


From a bird's-eye perspective, each DGM layer takes as an input the original mini-batch inputs $\vx$ (in our case this is the set of randomly sampled time-space points) and the output of the previous DGM layer. This process culminates with a vector-valued output $\vy$ which consists of the neural network approximation of the desired function $u$ evaluated at the mini-batch points. See \hyperref[fig:networkoutside]{Figure \ref{fig:networkoutside}} for a visualization of the overall architecture. \\

\vspace{0.5cm}

\begin{figure}[h!]
	\centering
	\scalebox{0.75}{
	\begin{tikzpicture}
	
	\definecolor{yellowd}{RGB}{255,187,51}
	\definecolor{yellowf}{RGB}{255,221,153}
	\definecolor{greend}{RGB}{102,255,102}
	\definecolor{greenf}{RGB}{153,255,153}
	\definecolor{blued}{RGB}{51,187,255}
	\definecolor{bluef}{RGB}{153,221,255}
	\definecolor{pinkd}{RGB}{210,121,164}
	\definecolor{pinkf}{RGB}{230,179,204}
	
	\tikzstyle{scircle}=[circle, thick, draw=greend, fill=greenf, minimum size=1cm]
	\tikzstyle{xcircle}=[circle, thick, draw=blued, fill=bluef, minimum size=1cm]
	\tikzstyle{ycircle}=[circle, thick, draw=pinkd, fill=pinkf, minimum size=1cm]
	\tikzstyle{smallrect}=[rectangle, thick, draw=yellowd, fill=yellowf, minimum height=1.5cm, minimum width=4.0cm, rounded corners]
	\tikzstyle{arrow}=[draw=black, -{Latex[length=2.5mm]}, thick]
	
	\node[] (SX1) {};
	\node[right=of SX1] (SX2) {};
	
	\node[xcircle, left=of SX1, yshift=1cm] (Sold) {$S^{old}$};
	\node[xcircle, left=of SX1, yshift=-1cm] (X) {$x$};
	
	\node[smallrect, right=of SX2] (GR) {$\vu^z \cdot \vx + \vw^z \cdot S + \vb^z$};
	\node[smallrect, above=of GR] (ZR) {$\vu^g \cdot \vx + \vw^g \cdot S + \vb^g$};
	\node[smallrect, below=of GR] (RR) {$\vu^r \cdot \vx + \vw^r \cdot S + \vb^h$};
	
	\node[scircle, right=of ZR] (Z) {$Z$};
	\node[scircle, right=of GR] (G) {$G$};
	\node[scircle, right=of RR] (R) {$R$};
	
	\node[smallrect, right=of Z] (SR) {$(1-G) \odot H + Z \odot S$};
	\node[smallrect, right=of R] (HR) {$\vu^h \cdot \vx + \vw^h \cdot (S \odot R) + \vb^h$};
	\node[scircle] (H) at ($(HR)!0.5!(SR)$) {$H$};
	\node[ycircle, right=of SR] (Snew) {$S^{new}$};

	\draw[thick] (Sold) -| (SX1.center);
	\draw[thick] (X) -| (SX1.center);
	\draw[arrow] (SX1.center) -- (GR);
	\draw[arrow] (SX2.center) |- (ZR);
	\draw[arrow] (SX2.center) |- coordinate (lRR) (RR);
	\draw[arrow] (ZR) -- node[above] {$\sigma$} (Z);
	\draw[arrow] (GR) -- node[above] {$\sigma$} (G);
	\draw[arrow] (RR) -- node[above] {$\sigma$} (R);
	\draw[arrow] (Z) -- (SR);
	\draw[arrow] (R) -- (HR);
	\draw[arrow] (HR) -- node[left] {$\sigma$} (H);
	\draw[arrow] (H) -- (SR);
	\draw[arrow] (G) -| ([xshift=-1.0cm]SR.south);
	\draw[arrow] (SR) -- (Snew);
	
	\node[above=of ZR] (aZR) {};
	\draw[thick] (Sold) |- (aZR.center);
	\draw[arrow] (aZR.center) -| (SR);
	\node[below=of RR] (bRR) {};
	\draw[thick] (lRR) |- (bRR.center);
	\draw[arrow] (bRR.center) -| (HR);

	\end{tikzpicture}
	}
	\caption{Operations within a single DGM layer.}
	\label{fig:dgmlayer3}
\end{figure}
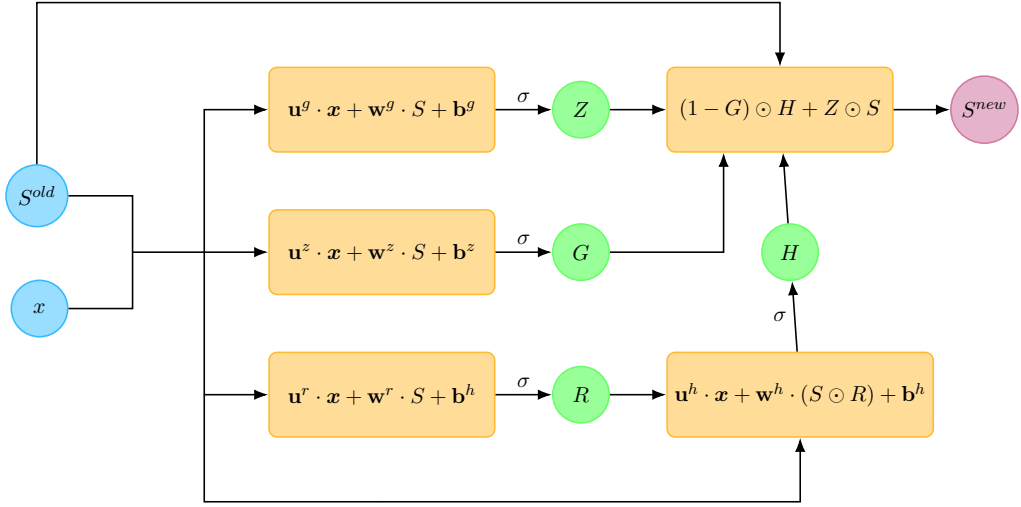

Within a DGM layer, the mini-batch inputs along with the output of the previous layer are transformed through a series of operations that closely resemble those in Highway Networks. Below, we present the architecture in the equations along with a visual representation of a single DGM layer in \hyperref[fig:dgmlayer3]{Figure \ref{fig:dgmlayer3}}:

\begin{align*}
S^1 ~&=~ \sigma \left( \vw^1 \cdot \vx + \vb^1 \right)
\\
Z^\ell ~&=~ \sigma \left(\vu^{z,\ell} \cdot \vx + \vw^{z,\ell} \cdot S^\ell + \vb^{z,\ell}  \right) && \ell = 1,..., L
\\
G^\ell ~&=~ \sigma \left(\vu^{g,\ell} \cdot \vx + \vw^{g,\ell} \cdot S^\ell + \vb^{g,\ell}  \right) && \ell = 1,..., L
\\
R^\ell ~&=~ \sigma \left(\vu^{r,\ell} \cdot \vx + \vw^{r,\ell} \cdot S^\ell + \vb^{r,\ell}  \right) && \ell = 1,..., L
\\
H^\ell ~&=~ \sigma \left(\vu^{h,\ell} \cdot \vx + \vw^{h,\ell} \cdot \left(S^\ell \odot R^\ell \right) + \vb^{h,\ell}  \right) && \ell = 1,..., L
\\
S^{\ell+1} ~&=~ \left(1 - G^\ell \right) \odot H^\ell + Z^\ell \odot S^\ell && \ell = 1,..., L \\
f(t,\vx;\vtheta) ~&=~ \vw \cdot S^{L+1} + \vb
\end{align*} 
where $\odot$ denotes Hadamard (element-wise) multiplication, $L$ is the total number of layers, $\sigma$ is an activation function and the $\vu$, $\vw$ and $\vb$ terms with various superscripts are the model parameters. \\

Similar to the intuition for LSTMs, each layer produces weights based on the last layer, determining how much of the information gets passed to the next layer. In \cite{sirignano2017dgm} the authors also argue that including repeated element-wise multiplication of nonlinear functions helps capture ``sharp turn'' features present in more complicated functions. Note that at every iteration the original input enters into the calculations of every intermediate step, thus decreasing the probability of vanishing gradients of the output function with respect to $x$. \\

Compared to a Multilayer Perceptron (MLP), the number of parameters in each hidden layer of the DGM network is roughly eight times bigger than the same number in an usual dense layer. Since each DGM network layer has 8 weight matrices and 4 bias vectors while the MLP network only has one weight matrix and one bias vector (assuming the matrix/vector sizes are similar to each other).	Thus, the DGM architecture, unlike a deep MLP, is able to handle issues of vanishing gradients, while being flexible enough to model complex functions.  \\

\emph{Remark on Hessian implementation:} second-order differential equations call for the computation of second derivatives. In principle, given a deep neural network $f(t,\vx; \vtheta)$, the computation of higher-order derivatives by automatic differentiation is possible. However, given $\vx \in \mathbb{R}^n$ for $n > 1$, the computation of those derivatives becomes computationally costly, due to the quadratic number of second derivative terms and the memory-inefficient manner in which the algorithm computes this quantity for larger mini-batches. For this reason, we implement a finite difference method for computing the Hessian along the lines of the methods discussed in \hyperref[chap:numerical]{Chapter \ref{chap:numerical}}. In particular, for each of the sample points $x$, we compute the value of the neural net and its gradients at the points $x + h e_j$ and $x - h e_j$, for each canonical vector $e_j$, where $h$ is the step size, and estimate the Hessian by central finite differences, resulting in a precision of order $\bigO(h^2)$. The resulting matrix $H$ is then symmetrized by the transformation $0.5 (H + H^T)$. 

%% file: ourWork.tex
\chapter{Implementation of the Deep Galerkin Method}
\setcounter{section}{6}

In this chapter we apply the Deep Galerkin Method to solve various PDEs that arise in financial contexts, as discussed in \hyperref[chap:PDE]{Chapter \ref{chap:PDE}}. The application of neural networks to the problem of numerically solving PDEs (and other problems) requires a great deal of experimentation and implementation decisions. Even
with the basic strategy of using the DGM method, there are already numerous decisions to make, including: 
\begin{itemize}
	\item the network architecture; 
	\item the size of the neural network to use to achieve a good balance between execution time and accuracy; 
	\item the choice of activation functions and other hyperparameters;
	\item the random sampling strategy, selection of optimization and numerical (e.g. differentiation and integration) algorithms, training intensity; 
	\item programming environment. \\
\end{itemize}

In light of this, our approach was to begin with simple and more manageable PDEs
and then, as stumbling blocks are gradually surpassed, move on to
more challenging ones. We present the results of applying DGM to the following problems:

\begin{enumerate}
	\item \emph{European Call Options}: \\
 	We begin with the Black-Scholes PDE, a \textbf{linear PDE} which has a simple analytical solution and is a workhorse model in finance. This also creates the basic setup for the remaining problems. 
	\item \emph{American Put Options}: \\
	Next, we tackle American options, whose main challenge is the \textbf{free boundary problem}, which needs to be found as part of the solution of the problem. This requires us to adapt the algorithm (particularly, the loss function) to handle this particular detail of the problem.
	\item \emph{The Fokker-Plank Equation}: \\
	Subsequently, we address the Fokker-Planck equation, whose solution is a probability
	density function that has special \textbf{constraints} (such as being
	positive on its domain and integrating to one) that need to met by the method.
	\item \emph{Stochastic Optimal Control Problems}: \\
	For even more demanding challenges, we focus on HJB equations,
	which can be highly \textbf{nonlinear}. In particular, we consider two optimal control problems: the Merton problem and the optimal execution problem.
	\item \emph{Systemic Risk}: \\ 
	The systemic risk problem allows us to apply the method to a \textbf{multidimensional system of HJB equations}, which involves multiple variables and equations with a high degree of nonlinearity. 
	\item \emph{Mean Field Games}: \\ 
	Lastly, we close our work with mean field games, which are formulated in terms of \textbf{conversant HJB and Fokker-Planck equations}. \\
\end{enumerate} 

The variety of problems we manage to successfully apply the method to attests to the power and flexibility of the DGM approach. \\

\subsection{How this chapter is organized}
Each section in this chapter highlights one of the case studies mentioned in the list above. We begin with the statement of the PDE and its analytical solution and proceed to present (possibly several) attempted numerical solutions based on the DGM approach. The presentation is done in such a way as to highlight the \textit{experiential} aspect of our implementation. As such the first solution we present is by no means the best, and the hope is to demonstrate the learning process surrounding the DGM and how our solutions improve along the way. Each example is intended to highlight a different challenge faced - usually associated with the difficulty of the problem, which is generally increasing in examples - and a proverbial ``moral of the story.'' \\

An important caveat is that, in some cases, we don't tackle the full problem in the sense that the PDEs that are given at the start of each  section are not always in their primal form. The reason for this is that the PDEs may be too complex to implement in the DGM framework directly. This is especially true with HJB equations which involve an optimization step as part of the first order condition. In these cases we resort to to simplified versions of the PDEs obtained using simplifying ansatzes, but we emphasize that even these can be of significant difficulty. \\

\textit{Remark (a note on implementation): }
in all the upcoming examples we use the same network architecture used by \cite{sirignano2017dgm} presented in \hyperref[chap:DGM]{Chapter \ref{chap:DGM}}, initializing the weights with Xavier initialization. The network was trained for a number of iterations (epochs) which may vary by example, with random re-sampling of points for the interior and terminal conditions every 10 iterations. We also experimented with regular dense feedforward neural networks and managed to have some success fitting the first problem (European options) but we found them to be less likely to fit more irregular functions and more unstable to hyperparameters changes as well. \\

\subsection{European Call Options}

\begin{mytheo}{One-Dimensional Black-Scholes PDE}{theoexample}
	\vspace{0.3cm}
	\begin{equation*}
	\begin{cases}
	\partial_t g(t,x) + r x \cdot \partial_x g(t,x)+ \frac{1}{2} \sigma^2 x^2\cdot \partial_{xx} g(t,x)  ~=~  r \cdot g(t,x) 
	\\
	g(T,x) ~=~ G(x)
	\end{cases}
	\end{equation*} \\ 
	\vspace{0.3cm}
	\underline{Solution:}
	\vspace{-0.4cm}
	\begin{align*}
	g(t,x) ~&=~ x ~ \Phi(d_+) - K e^{-r(T-t)} \Phi(d_-) \\
	\mbox{where } \quad d_{\pm} ~&=~ \frac{\ln(x/K) + (r \pm \frac{1}{2} \sigma^2) (T-t) }{\sigma \sqrt{T-t}}
	\end{align*}  	
\end{mytheo} 
\vspace{0.5cm}

\begin{figure}[t!]
	\centering
	\includegraphics[width=0.6\linewidth]{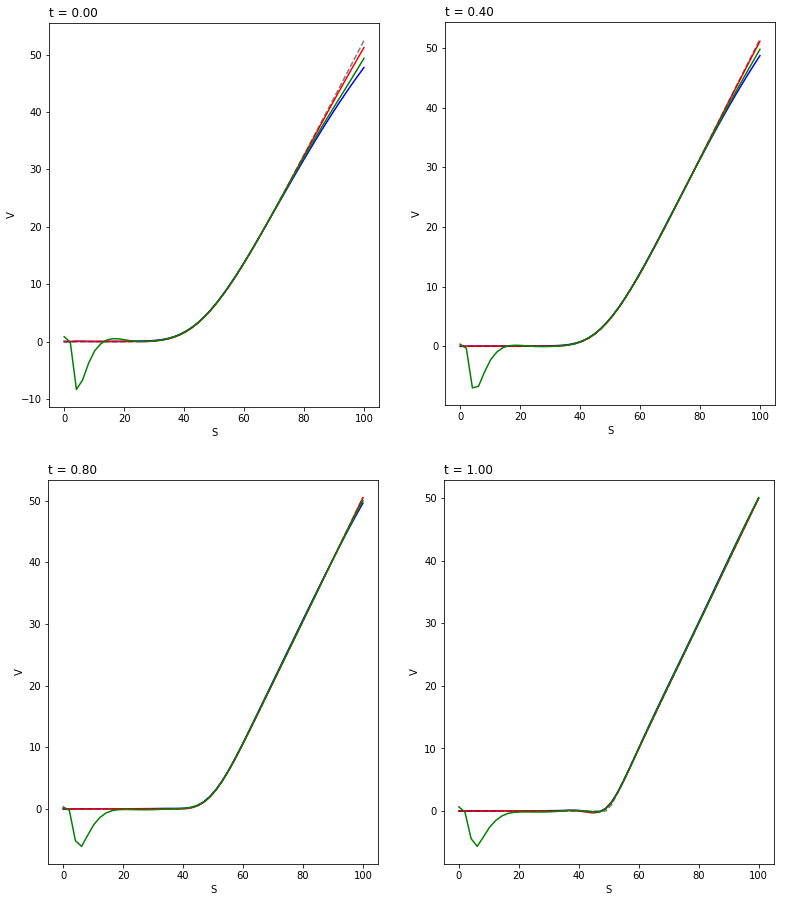}
	\caption{Call Prices as a function of Stock Price: the black dashed line is the true value function, calculated using the Black and Scholes Formula; the green, blue and red lines correspond to the three sampling methodologies described above.}
	\label{fig:call-at-diff-maturities}
\end{figure}

As a first example of the DGM approach, we trained the network to learn the value of a European call option. For our experiment, we used the interest rate $r = 5\%$, the volatility $\sigma = 25\%$, the initial stock price $S_0 = 50$, the maturity time $T=1$ and the option's strike price $K=50$. In \hyperref[fig:call-at-diff-maturities]{Figure \ref{fig:call-at-diff-maturities}} we present the true and estimated option values at different times to maturity. \\

\begin{figure}[h]
	\centering
	\includegraphics[width=0.7\linewidth]{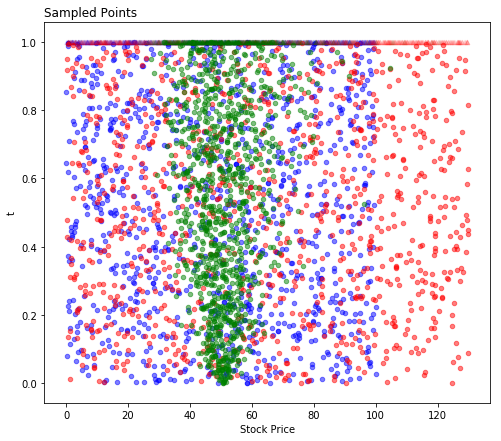}
	\caption{Different sampling schemes: lognormal (green), uniform on $[0,1] \times [0,100]$ (blue) and uniform on $[0,1] \times [0,130]$ (red)}
	\label{fig:sampled-points}
\end{figure}

First, we sampled uniformly on the time domain and according to a lognormal distribution on the space domain as this is the exact distribution that the stock prices follow in this model. We also sampled uniformly at the terminal time point. However, we found that this did not yield good results for the estimated function. These sampled points and fits can be seen in the green dots and lines in \hyperref[fig:sampled-points]{Figure \ref{fig:sampled-points}} and \hyperref[fig:call-at-diff-maturities]{Figure \ref{fig:call-at-diff-maturities}}. 

Since the issues seemed to appear at regions that were not well-sampled we returned to the approach of \cite{sirignano2017dgm} and sampled uniformly in the region of interest $[0,1] \times [0,100]$. This improved the fit, as can be seen in the blue lines of \hyperref[fig:call-at-diff-maturities]{Figure \ref{fig:call-at-diff-maturities}}, however, there were still issues on the right end of the plots with the fitted solution dipping too early. \\

Finally, we sampled uniformly \textit{beyond} the region of interest on $[0,1] \times [0,130]$ to show the DGM network points that lie to the right of the region of interest. This produced the best fit, as can be seen by the red lines in \hyperref[fig:call-at-diff-maturities]{Figure \ref{fig:call-at-diff-maturities}}. \\

Another point worth noting is that the errors are smaller for times that are closer to maturity. This reason for this behavior could be due to the fact that the estimation process is ``drawing information'' from the terminal condition. Since this term is both explicitly penalized and heavily sampled from, this causes the estimated function to behave well in this region. As we move away from this time point, this stabilizing effect diminishes leading to increased errors. \\

\underline{\Large \textit{\textbf{Moral: sampling methodology matter!}}}  \\


%
%

\subsection{American Put Options}

\begin{mytheo}{Black-Scholes PDE with Free Boundary}{theoexample}
	\vspace{0.0cm}
	\begin{equation*}
		\begin{cases}
		\partial_t g + r x \cdot \partial_x g + \frac{1}{2} \sigma^2 x^2\cdot \partial_{xx} g - r \cdot g = 0 & \qquad \left\{ (t,x): g(t,x) > G(x) \right\} 
		\\
		g(t,x) \geq G(x) & \qquad (t,x) \in [0,T] \times \RR 
		\\
		g(T,x) ~=~ G(x) &  \qquad x \in \RR
		\end{cases}
	\end{equation*}
	where $G(x) = (K-x)_+$ \\ 
	
	\underline{Solution:} ~~~ No analytical solution.	
\end{mytheo} 
\vspace{0.5cm}

In order to further test the capabilities of DGM nets, we trained the network to learn the value of American-style put options. This is a step towards increased complexity, compared to the European variant, as the American option PDE formulation includes a free boundary condition. We utilize the same parameters as the European call option case: $r = 5\%$, $\sigma = 25\%$, $S_0 = 50$, $T=1$ and $K=50$. \\

In our first attempt, we trained the network using the method prescribed by \cite{sirignano2017dgm}. The approach for solving free boundary problems there is to sample uniformly over the region of interest ($t \in [0,1], S \in [0,100]$ in our case), and accept/reject training examples for that particular batch of points, depending on whether or not they are inside or outside the boundary region implied by the last iteration of training. This approach was able to correctly recover option values. \\

As an alternative approach, we used a different formulation of the loss function that takes into account the free boundary condition instead of the acceptance/rejection methodology. In particular, we applied a loss to all points that violated the condition $\left\{ (t,x): g(t,x) \geq G(x) \right\}$ via: 
\[ \bigg\| \max\{ - \left(f(t,x;\vtheta) - (K-x)_+\right), 0 \} \bigg\|^2_{[0,T] \times \Omega, ~\nu_1} \] 

\begin{figure}[h!]
	\centering
	\includegraphics[width=0.63\linewidth]{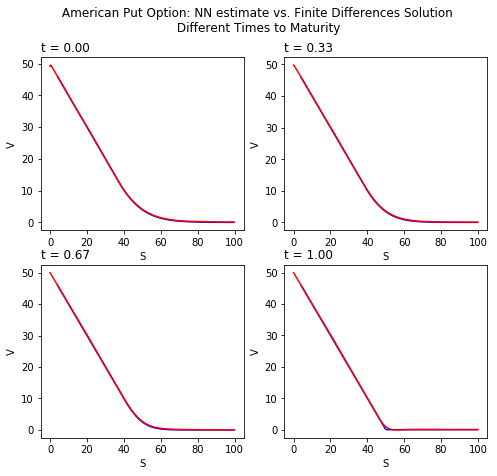}
	\caption{Comparison of American put option prices at various maturities computed using DGM (red) vs. finite difference methods (blue)}
	\label{fig:americanputoptionnnestimatevsfinitedifferencessolution}
\end{figure}

\hyperref[fig:americanputoptionnnestimatevsfinitedifferencessolution]{Figure \ref{fig:americanputoptionnnestimatevsfinitedifferencessolution} } compares the DGM fitted option prices obtained using the alternative inequality loss for different maturities compared to the finite difference method approach. The figure shows that we are successful at replicating the option prices with this loss function. \\

\begin{figure}[h!]
	\centering
	\includegraphics[width=0.6\linewidth]{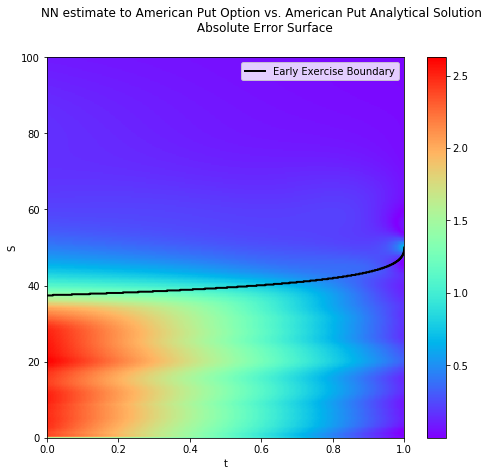}
	\caption{Absolute difference between DGM-estimated American put option prices and analytical solution for corresponding European put options.}
	\label{fig:comparisonamericancallvanalyticalabserrorplotsurfaceandboundary}
\end{figure}

\hyperref[fig:comparisonamericancallvanalyticalabserrorplotsurfaceandboundary]{Figure \ref{fig:comparisonamericancallvanalyticalabserrorplotsurfaceandboundary} }
depicts the absolute error between the estimated put option values and the analytical prices for corresponding European puts given by the Black-Scholes formula. Since the two should be equal in the continuation region, this can be a way of indirectly obtaining the early exercise boundary. The black line is the boundary obtained by the finite difference method  and we see that is it closely matched by our implied exercise boundary. The decrease in the difference between the two option prices below the boundary as time passes reflects the deterioration of the early exercise optionality in the American option.  \\

\vspace{0.5cm}

\underline{\Large \textit{\textbf{Moral: loss functions matter!}}}  \\

\subsection{Fokker-Planck Equations} \label{sec:FK}

\begin{mytheo}{Fokker-Planck Equation for OU process with random Gaussian start}{theoexample}
	\begin{equation*}
	\begin{cases}
	~ \partial_t p + \kappa \cdot p + \kappa(x - \theta) \cdot \partial_x p - \frac{1}{2} \sigma^2 \cdot \partial_{xx} p  = 0 \qquad \quad  (t,x) \in \RR_+ \times \RR
	\\
	~ p(0,x) = \frac{1}{\sqrt{2 \pi v} } \cdot e^{-\tfrac{x^2}{2 v} }
	\end{cases}
	\end{equation*}
	\underline{Solution:} ~~~ Gaussian density function.	
\end{mytheo} 
\vspace{0.5cm}

The Fokker-Planck equation introduces a new difficulty in the form of a constraint on the solution. We applied the DGM method to the Fokker-Planck equation for the Ornstein\textendash Uhlenbeck mean-reverting process. If the process begins at a fixed point $x_0$, i.e. its
initial distribution is a Dirac delta at $x_0$, then the
solution for this PDE is known to have the normal distribution
\[X_{T}~\sim~N\left(x_{0} \cdot e^{-\kappa(T-t)}+\theta\left(1-e^{-\kappa(T-t)}\right),~\frac{\sigma^{2}}{2\kappa}\left(1-e^{-2\kappa(T-t)}\right)\right)
\]
Since it is not possible to directly represent the initial delta numerically, one would have to approximate it, e.g. with a normal distribution with mean $x_0$ and a small variance. In the case where the starting point is Gaussian, we use Monte Carlo simulation to determine the distribution at every point in time, but we note that the distribution should be Gaussian since we are essentially using a conjugate prior. \\ 

\begin{figure}[h!]
	\centering
	\includegraphics[width=0.7\linewidth]{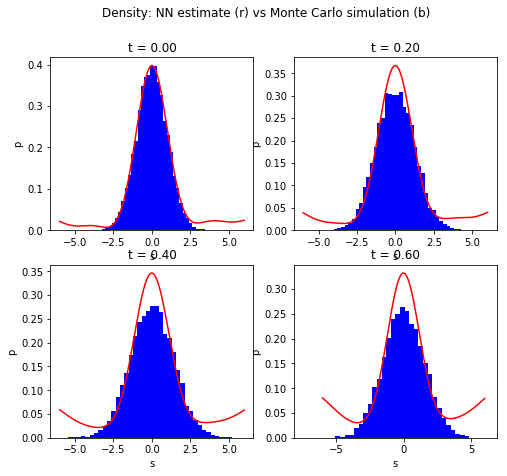}
	\caption{Distribution of $X_t$ at different times. Blue bars correspond to histograms of simulated values; red lines correspond to the DGM solution of the required Fokker-Planck equation.}
	\label{fig:fokkerplanck1sttry}
\end{figure}

For the DGM algorithm, we used loss function terms for the differential
equation itself, the initial condition and added a penalty to reflect the non-negativity constraint. Though we intended to include another term  to force the integral of the solution to equal one, this
proved to be too computationally expensive, since an integral must be numerically evaluated at each step of the network training phase. For the parameters $\theta=0.5$, $\sigma=2$, $T=1$, $\kappa = 0$, \hyperref[fig:fokkerplanck1sttry]{Figure \ref{fig:fokkerplanck1sttry} }  shows the density estimate $p$ as a function of position $x$ at different time points compared to the simulated distribution. As can be seen from these figures, the fitted distributions had issues around the tails and with the overall height of the fitted curve, i.e. the fitted densities did not integrate to 1. The neural network estimate, while correctly approximating the initial condition, is not able to conserve probability mass and the Gaussian bell shape across time. \\

To improve on the results, we apply a change of variables:
\[p\left(t,x\right)=\frac{e^{-u\left(t,x\right)}}{c(t)}\]
where $c(t)$ is a normalizing constant. This amounts to fitting an exponentiated normalized neural network guaranteed to remain positive and integrate to unity. This approach provides an alternative PDE to be solved by the DGM method:
\[\partial_t u +\kappa(x-\theta)\partial_{x}u - \dfrac{\sigma^2}{2}\left[\partial_{xx}u - (\partial_{x}u)^2 \right] = \kappa 
  + \frac{\int (\partial_t u) e^{-u} dx}{\int e^{-u} dx} \]

Notice that the new equation is a non-linear PDE dependent on a integral term. To handle the integral term and avoid the costly operation of numerically integrating at each step, we first uniformly sample $\{t_j\}_{j=1}^{N_t}$ from $t \in [0,T]$ and $\{x_k\}_{q=1}^{N_x}$ from $[x_{min},x_{max}]$, then, for each $t_j$, 
we use \textbf{importance sampling} to approximate the expectation term by 
\[ I_t := \sum_{k=1}^{N_x} (\partial_t u(t_j,x_k)) w(x_k) \] 
where 
\[ w(x_k) = \tfrac{e^{u(t_j,x_k)}}{\sum_{k=1}^{N_x} e^{u(t_j,x_k)} } \] 
Note that since the density for uniform distribution is constant within the sampling region, 
the denominator terms for the weights are cancelled. The $L_1$ loss is then approximated by:
\begin{equation*}
\frac{1}{N_t} \frac{1}{N_x} \sum_{j=1}^{N_t}  \sum_{k=1}^{N_x} (\partial_t + \mathcal{L}) u(t_j,x_k,I_t,\theta)
\end{equation*}

\begin{figure}[h!]
	\centering
	\includegraphics[width=0.75\linewidth]{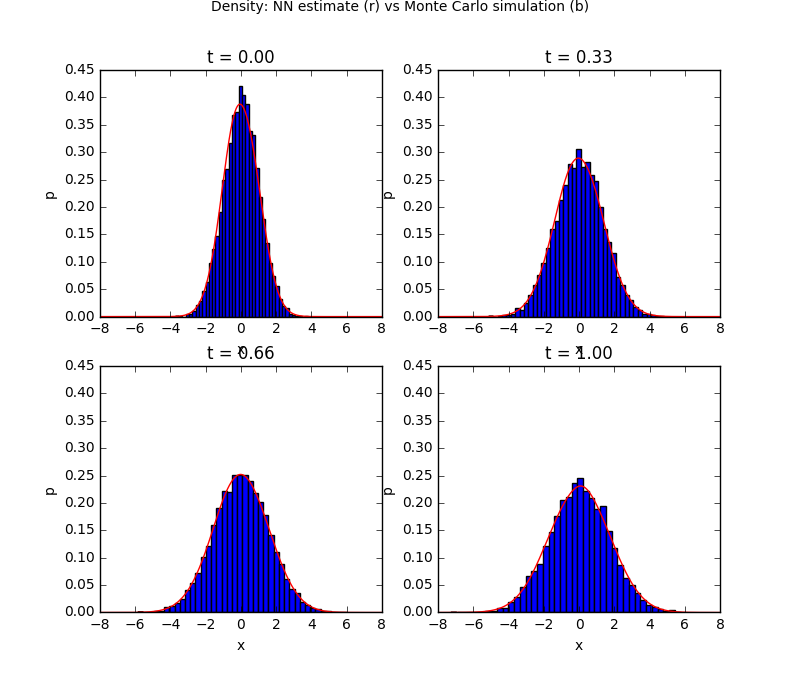}
	\caption{{Distribution of $X_t$ at different times. Blue bars correspond to histograms of simulated values; red lines correspond to the DGM solution of the required Fokker-Planck equation using the modified approach.}}
	\label{fig:fokkerplanck2ndtry}
\end{figure}

Even though the resulting equation turns out to be more complex, using this technique to train the network by solving for $u(x,t)$ and transforming back to $p(x,t)$ allowed us to achieve stronger results as evidence by the plots in \hyperref[fig:fokkerplanck2ndtry]{Figure \ref{fig:fokkerplanck2ndtry} }.

\vspace{0.3cm}

Notice that the network was able to accurately recover the shape and preserve the probability mass across time steps. \\

It is interesting to note that, in this example, the loss of linearity in the PDE was not as important to being able to solve the problem than imposing the appropriate structure on the desired function. \\

\vspace{0.7cm}

\underline{\Large \textit{\textbf{Moral: prior knowledge matters!}}} \\

\subsection{Stochastic Optimal Control Problems}

In this section we tackle a pair of nonlinear HJB equations. The interest is in both the value function as well as the optimal control. The original form of the HJB equations contains an optimization term (first-order condition) which can be difficult to work with. Here we are working with the simplified PDE once the optimal control in feedback form is substituted back in and an ansatz is potentially used to simplify further. Since we are interested in both the value function and the optimal control, and the optimal control is written in terms of derivatives of the value function, a further step of numerically differentiating the DGM output (based on finite differences) is required for the optimal control. \\

\subsubsection{Merton problem}

\begin{mytheo}{Merton Problem - Optimal Investment with Exponential Utility}{theoexample}
	\begin{equation*}
	\vspace{0.5cm}
	\begin{cases}
	~ \partial_t H - \frac{\lambda^2}{2\sigma^2} \frac{(\partial_x H)^2}{\partial_{xx} H} + r x H = 0 \qquad \quad  (t,x) \in \RR_+ \times \RR
	\\
	~ H(T,q) = -\alpha q^2 
	\end{cases}
	\end{equation*}
	\vspace{0.5cm}	
	\underline{Solution (value function and optimal control):}
	\vspace{-0.4cm}
	\begin{align*}
	H(t,x)  &= - \exp \left[ {-x \gamma e^{r(T-t)} - \tfrac{\lambda^2}{2}  (T-t)	} \right] 
	\\
	\pi^*_t &= \frac{\lambda}{\gamma \sigma} e^{-r(T-t)}
	\\
	\mbox{where } \qquad \lambda &= \frac{\mu - r}{\sigma}
	\end{align*}		
\end{mytheo} 
\vspace{0.5cm}

In this section, we attempt to solve the HJB equation for the Merton problem with exponential utility. In our first attempts, we found that the second-order derivative appearing in the denominator in the above equation generates large instabilities in the numerical resolution of the problem. Thus, we rewrite the equation by multiplying out to obtain:

\begin{equation*}
- \frac{\lambda^2}{2\sigma^2} (\partial_x H)^2 + \partial_{xx} H \left( \partial_t H - \frac{\lambda^2}{2 \sigma^2}  + r x H \right) = 0
\end{equation*}

In this formulation, the equation becomes a quasi-linear PDE which was more stable numerically. The equation was solved with  parameters $r = 0.05$, $\sigma = 0.25$, $\mu = 0.2$ and $\gamma = 1$, 
with terminal time $T = 1$, in the region $(t,x) \in [0,1]^2$, with oversampling of $50\%$ in the $x$-axis. 
\begin{figure}[H]
	\centering
	\includegraphics[width=0.9\textwidth]{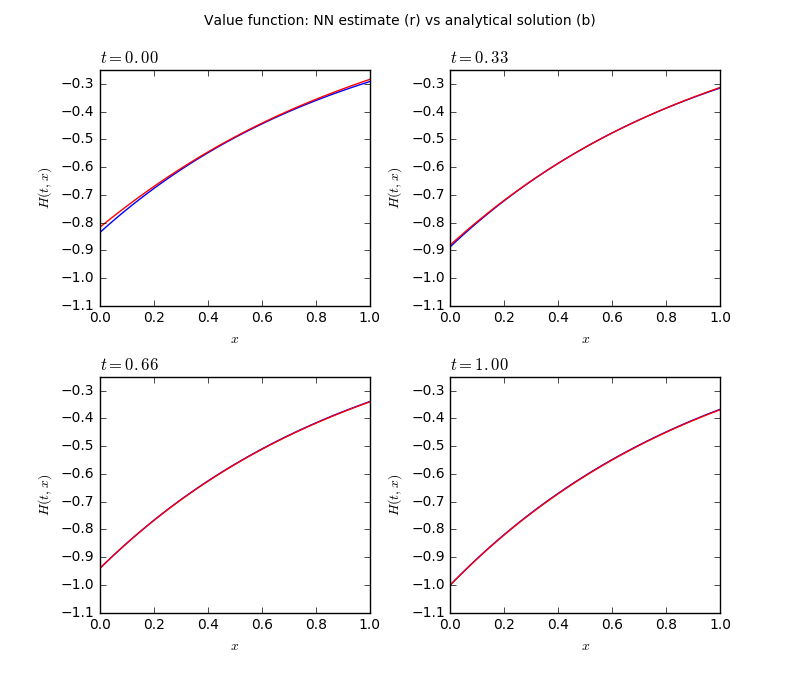}
	\caption{Approximate (red) vs. analytical (blue) value function for the Merton problem.}
	\label{fig:merton_value_function}
\end{figure}

\begin{figure}[H]
	\centering
	\includegraphics[width=0.49\linewidth]{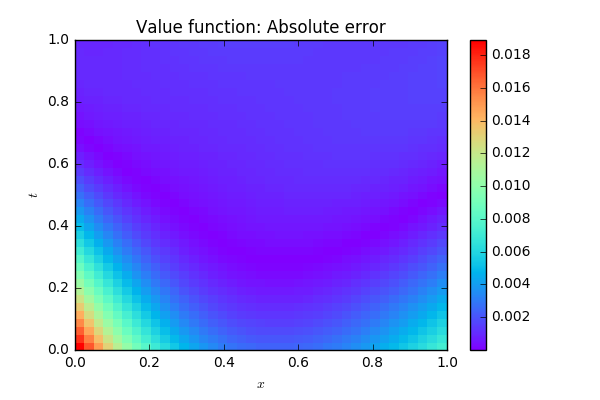} 
	\includegraphics[width=0.49\linewidth]{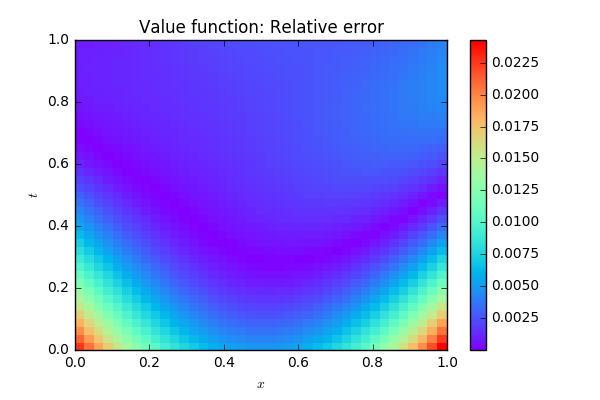}	
	\caption{Absolute (left panel) and relative (right panel) error between approximate and analytical solutions of the Merton problem value function.}
	\label{fig:merton_value_function_err}
\end{figure}

\begin{figure}[H]
	\centering
	\includegraphics[width=0.9\textwidth]{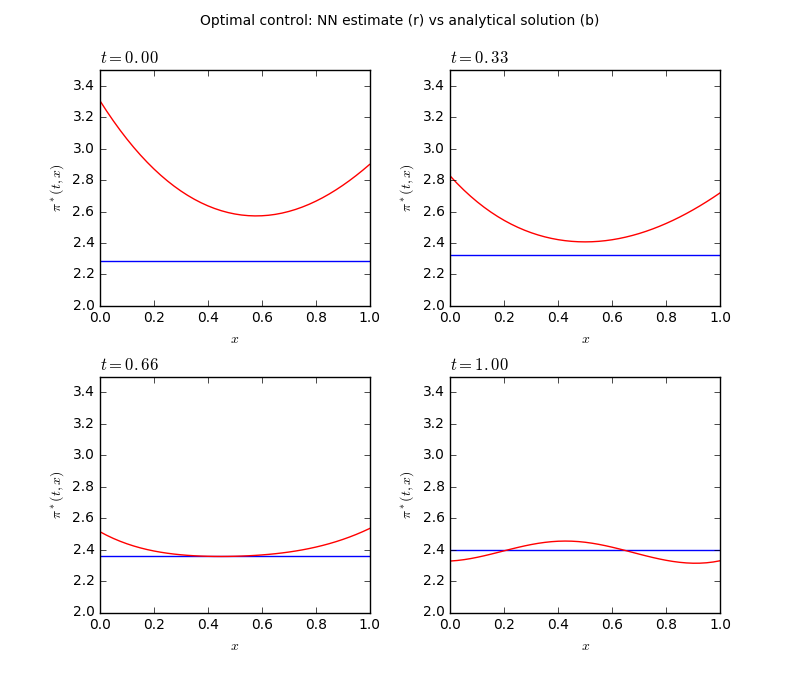}
	\caption{Approximate (red) vs. analytical (blue) optimal control for the Merton problem.}
	\label{fig:merton_opt_control}
\end{figure}

\begin{figure}[H]
	\centering
	\includegraphics[width=0.49\linewidth]{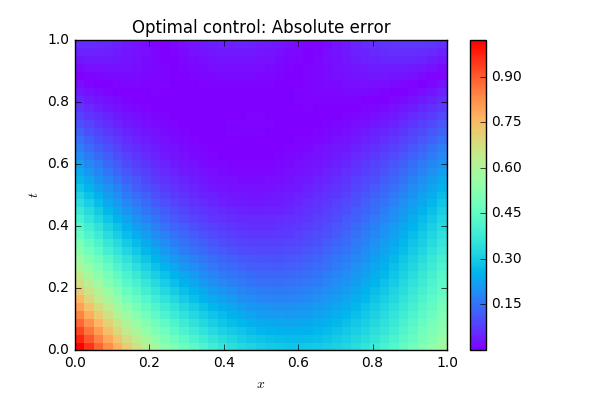}
	\includegraphics[width=0.49\linewidth]{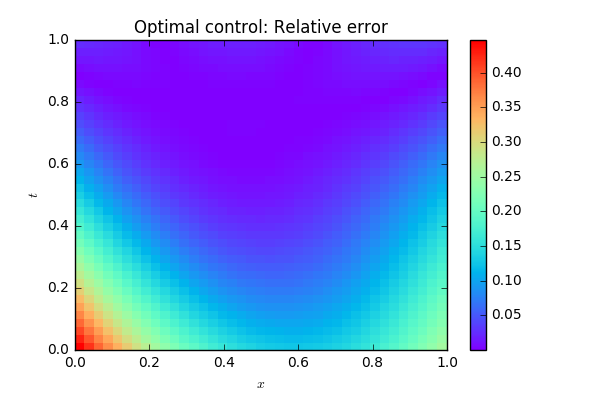}
	\caption{Absolute (left panel) and relative (right panel) error between approximate and analytical solutions of the optimal control.}
	\label{fig:merton_value_control_err}
\end{figure}

\clearpage

The estimated value function (\hyperref[fig:merton_value_function]{Figures  \ref{fig:merton_value_function}} and \ref{fig:merton_value_function_err}) and optimal control (\hyperref[fig:merton_opt_control]{Figure \ref{fig:merton_opt_control}}) are compared with the analytical solution below. Examining the plots, we find that the value function is estimated well by the neural network. Notice, however that at $t = 0$, the error between the approximate and analytical solutions is larger, but within an acceptable range. This may once again be due to the fact that the terminal condition has a stabilizing effect on the solution that diminished as we move away from that time point. \\

In general, we are interested in the optimal control associated with the HJB equation. In this context, the optimal control involves dividing by the second-order derivative of the value function which appears to be small in certain regions. This causes a propagation of errors in the computed solution, as seen in \hyperref[fig:merton_opt_control]{Figures \ref{fig:merton_opt_control} and \ref{fig:merton_value_control_err}}. The approximation appears to be reasonably close at $t = 1$, but diverges quickly as $t$ goes to $0$. Notice that regions with small errors in the value function solution correspond to large errors in the optimal control. \\

\subsubsection{Optimal Execution}

\begin{mytheo}{Optimal Liquidation with Permanent and Temporary Price Impact}{theoexample}
	\begin{equation*}
	\vspace{0.5cm}
	\begin{cases}
	~ \partial_t h(t,q) - \phi q^2  + \frac{1}{4 \kappa} \left( bq + \partial_q h(t,q) \right)^2 = 0 \qquad \quad  (t,q) \in \RR_+ \times \RR
	\\
	~ h(T,q) = -\alpha q^2 
	\end{cases}
	\end{equation*}
	\vspace{0.5cm}	
	\underline{Solution:}
	\vspace{-0.6cm}
	\begin{align*}
	h(t) &= \sqrt{k \phi} \cdot \frac{1 + \zeta e^{2\gamma(T-t)}}{1 - \zeta e^{2\gamma(T-t)}} \cdot q^2
	\\
	\mbox{where } \qquad 
	\gamma &= \sqrt{\frac{\phi}{k}} , \qquad 
	\zeta = \frac{\alpha - \tfrac{1}{2}b + \sqrt{k \phi}}{\alpha - \tfrac{1}{2}b - \sqrt{k \phi}}
	\end{align*}  		
\end{mytheo} 
\vspace{0.5cm}

For the second nonlinear HJB equation, the optimal execution problem was solved with parameters with $k = 0.01$, 
$b = 0.001$, $\phi = 0.1$, $\alpha = 0.1$, from $t = 0$ to terminal time $t = T = 1$, 
with $q \in [0,5]$, with oversampling of $50 \%$ in the $q$-axis. The approximation in the plots below shows a good fit to the true value function. The optimal control solution for the equation only depends on the  first derivative of the solution, so the error propagation is not as large as in the previous problem, as shown in the computed solution, 
where there is a good fit, worsening when $q$ goes to $0$ and $t$ goes to $T$.

\begin{figure}[h!]
	\centering
	\includegraphics[width=0.9\textwidth]{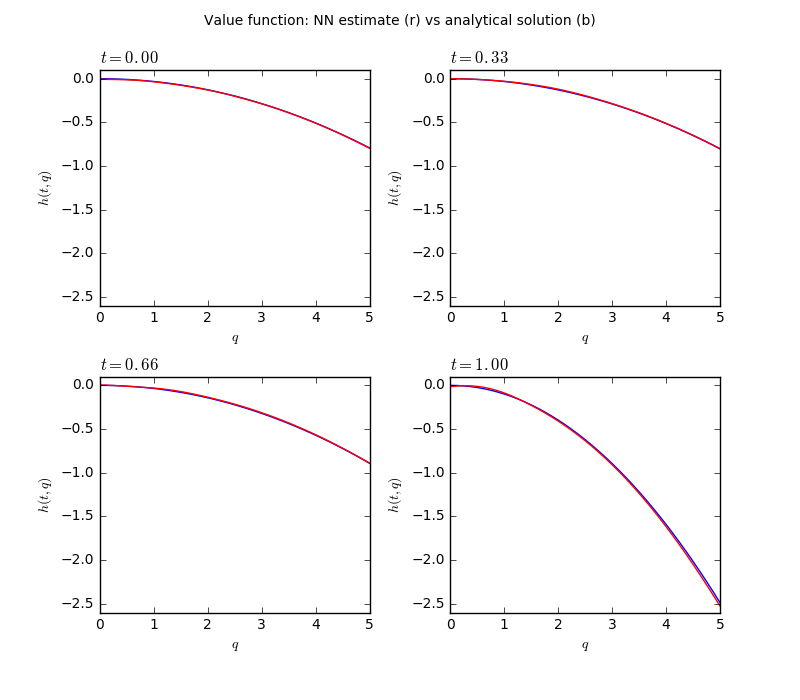}
	\caption{Approximate (red) vs. true (blue) value function for the optimal execution problem.}
\end{figure}
\begin{figure}[H]
	\centering
	\includegraphics[width=0.65\textwidth]{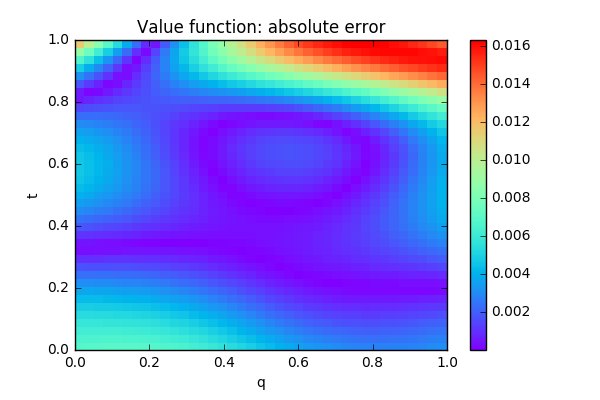}
	\caption{Absolute error between approximate and analytical solutions for
		the value function of optimal execution problem.}
\end{figure}
\begin{figure}[h!]
	\centering
	\includegraphics[width=0.85\textwidth]{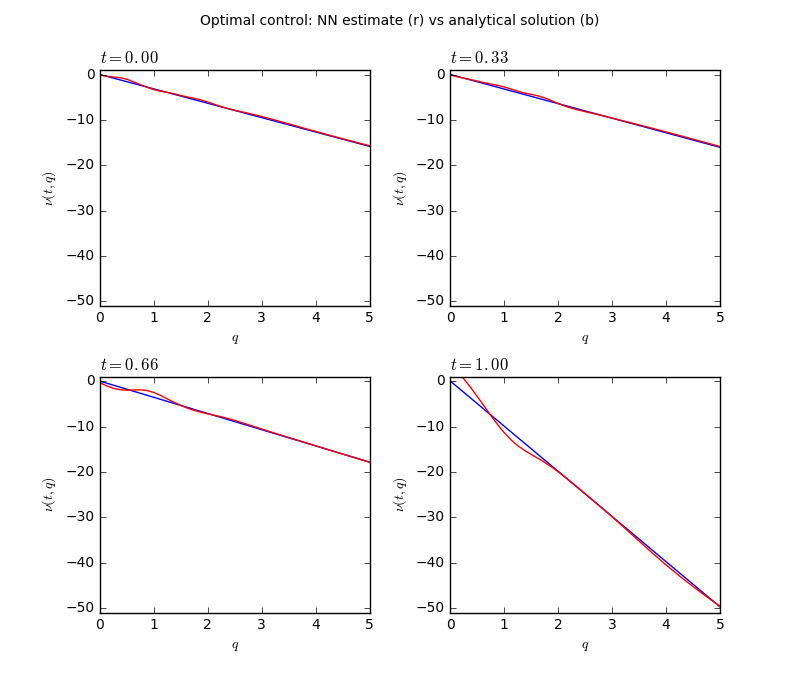}
	\caption{Approximate (red) vs. true (blue) optimal trading rate for the optimal execution problem.}
\end{figure}
\begin{figure}[h!]
	\centering
	\includegraphics[width=0.6\textwidth]{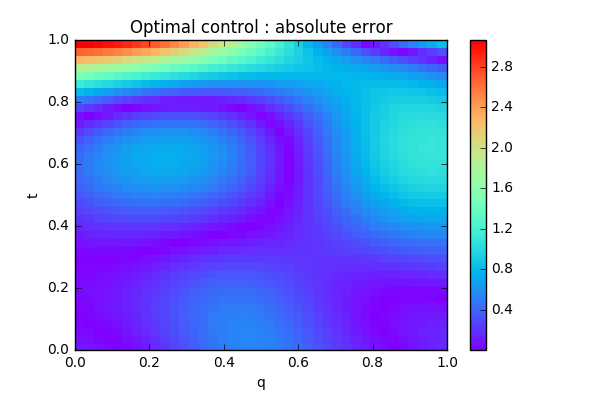}
	\caption{Absolute error between approximate and analytical solutions  
		for the optimal control in optimal execution problem.}
\end{figure}

\vspace{0.3cm}

\underline{\Large \textit{\textbf{Moral: going from value function to optimal control is nontrivial!}}}

\subsection{Systemic Risk}

\begin{mytheo}{Systemic Risk}{theoexample}
	\begin{equation*}
	\begin{cases}
	{\displaystyle
		\partial_t V^i + \sum_{j=1}^{N} \left[a(\overline{x} - x^j) - \partial_j V^j \right] \partial_j V^i }
	\\
	{ \displaystyle \hspace{1.5cm} + \frac{\sigma^2}{2} \sum_{j,k=1}^{N} \left( \rho^2 + \delta_{jk} (1-\rho^2) \right) \partial_{jk} V^i }
	\\
	{ \displaystyle \hspace{1.5cm} + \tfrac{1}{2} (\epsilon - q)^2 \left(\overline{x} - x^i \right)^2 + \tfrac{1}{2} \left( \partial_i V^i \right)^2 = 0 }
	\\
	{ \displaystyle V^i(T,\vx) = \tfrac{c}{2}  \left(\overline{x} - x^i \right)^2 \qquad \qquad  \qquad } & ~~ \mbox{for } ~~ i = 1, ..., N. 
	\end{cases}
	\end{equation*}
	
	\vspace{0.5cm}	
	\underline{Solution:}
	\vspace{-0.3cm}
	\begin{align*}
	V^i(t,\vx) &= \frac{\eta(t)}{2} \left(\overline{x} - x^i \right)^2 + \mu(t) 
	\\
	\alpha_t^{i,*} &=  \bigg( q + \left(1 - \tfrac{1}{N} \right) \cdot \eta(t) \bigg) \left( \overline{X}_t - X^i_t \right)
	\\
	\mbox{where } \qquad  \eta(t) &= \frac{ -(\epsilon - q)^2 \left( e^{(\delta^+ - \delta^-)(T-t)} - 1 \right) - c \left( \delta^+ e^{(\delta^+ - \delta^-)(T-t)} - \delta^- \right)}{ \left( \delta^-e^{(\delta^+ - \delta^-)(T-t)} - \delta^+ \right) - c (1 - \tfrac{1}{N^2}) \left( e^{(\delta^+ - \delta^-)(T-t)} - 1 \right)}
	\\ \mu(t) &= \tfrac{1}{2} \sigma^2 (1-\rho^2) \left( 1 - \tfrac{1}{N}  \right) \int_t^T \eta(s) ~ds
	\\ 
	\delta^{\pm} &= - (a+q) \pm \sqrt{R}, 
	\qquad \qquad 
	R = (a+q)^2 + \Big( 1 - \tfrac{1}{N^2} \Big)(\epsilon - q^2) \\
	\end{align*} 		
\end{mytheo} 
\vspace{0.3cm}

The systemic risk problems brings our first system of HJB equations (which happen to also be nonlinear). This was solved for the two-player ($N = 2$) case 
with correlation $\rho = 0.5$, $\sigma = 0.1$, $a = 1$, 
$q = 1$, $\epsilon = 10$, $c = 1$, from $t=0$ to terminal 
time $t = T = 1$, with $(x_1,x_2) \in [0,10] \times [0,10]$, 
and results were compared with the analytical solution. \\

Note that the analytical solution has two symmetries, one between the value functions for both players, and one around the ${x_1 = x_2}$ line. The neural network solution captures both symmetries, fitting the analytical solution for this system.  The regions with the largest errors were found in the symmetry axis, as $t$ goes to $0$, but away from those regions the error in the solution becomes very low. Once again this may be attributed to the influence of the terminal condition.

\begin{figure}[H]
	\centering
	\includegraphics[width=0.7\textwidth]{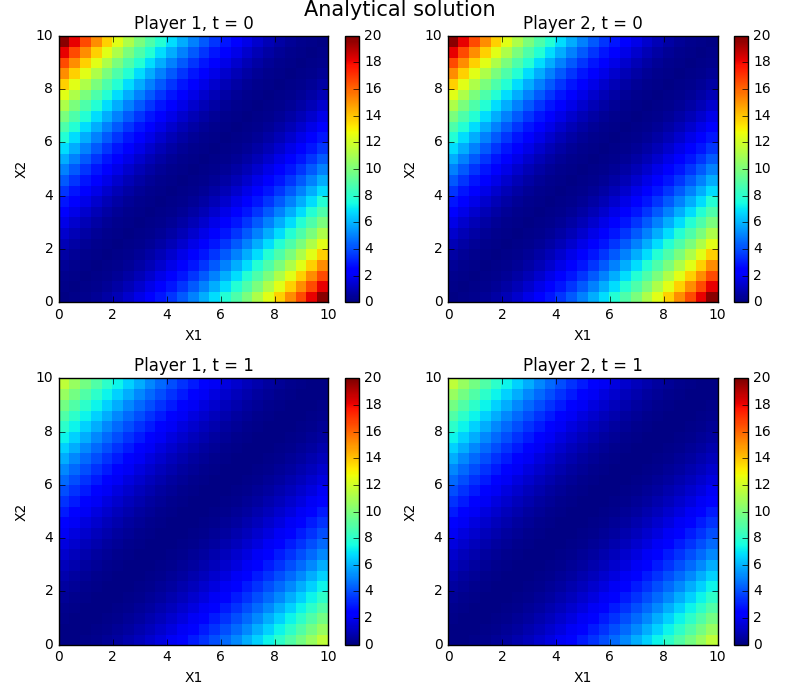}
	\caption{Analytical solution to the systemic risk problem.}
\end{figure}
\begin{figure}[H]
	\centering
	\includegraphics[width=0.7\textwidth]{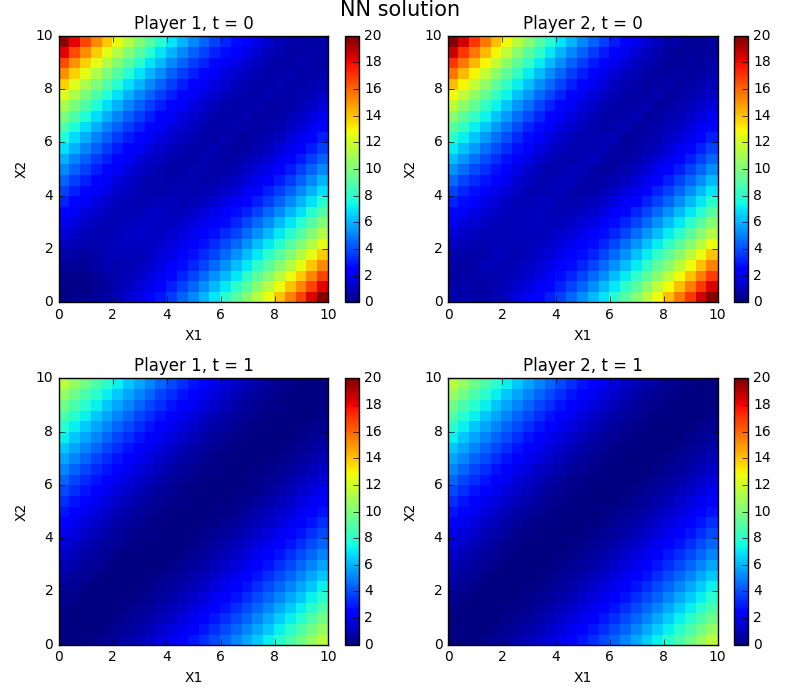}
	\caption{Neural network solution to the systemic risk problem.}
\end{figure}

\begin{figure}[H]
	\centering
	\includegraphics[width=0.7\linewidth]{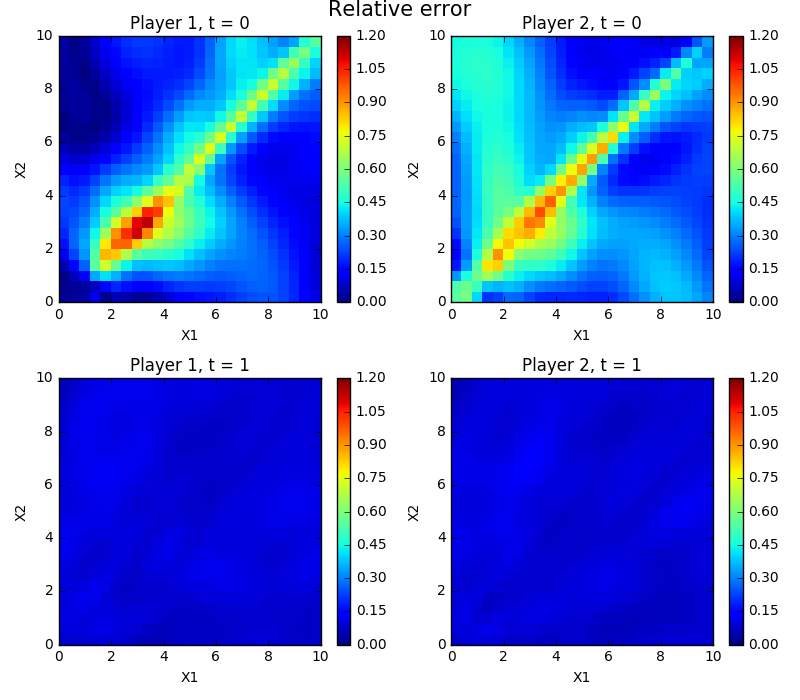}
	\caption{Absolute error between approximate and analytical solutions for the systemic risk problem.}
\end{figure}	
\begin{figure}[H]
	\centering
	\includegraphics[width=0.7\linewidth]{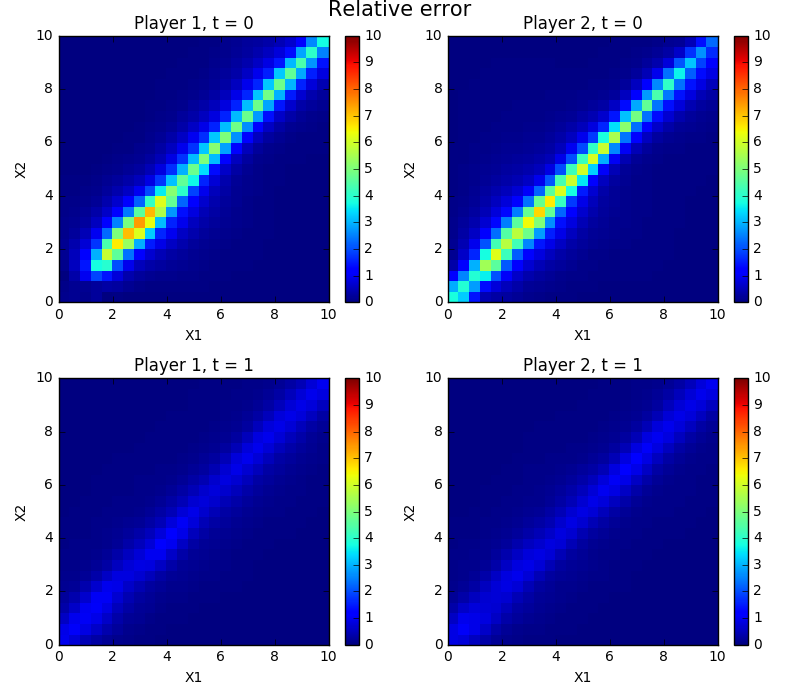}
	\caption{Relative error between approximate and analytical solutions for the systemic risk problem.}
\end{figure}

The systemic risk problem was also solved for five players with the same parameters as above to test the method's capability with higher dimensionality both in terms of the number of variables and the number of equations in the system. In the figures below, we compare the value function for a player when he deviates by $\Delta x$ from the initial state from $x_0$, with $x_0 = 5$. Note that all players have the same value function by symmetry. The plots show that the neural network trained using the DGM approach is beginning to capture the overall shape of the solution, although there is still a fair amount of deviation from the analytical solution. This suggests that more training time, or a better training procedure, should eventually capture the true solution with some degree of accuracy.

\begin{figure}[H]
	\centering
	\includegraphics[width=0.68\textwidth]{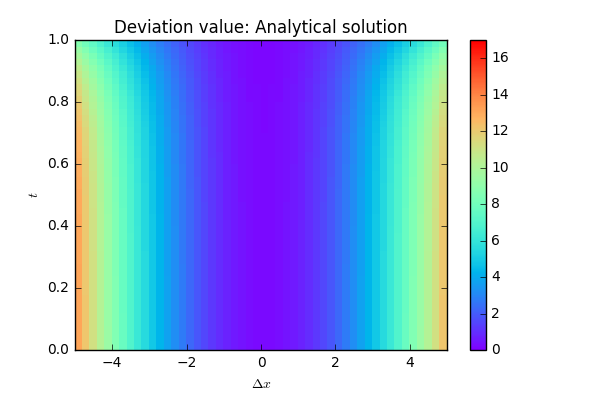}
	\caption{Analytical solution to the systemic risk problem with five players.}
\end{figure}
\begin{figure}[H]
	\centering
	\includegraphics[width=0.68\textwidth]{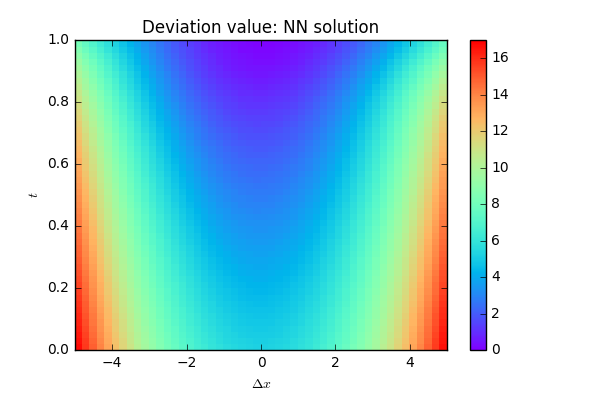}
	\caption{Neural network solution to the systemic risk problem with five players.}
\end{figure}

\subsection{Mean Field Games}

\begin{mytheo}{Optimal Liquidation in a Mean Field with Identical Preferences}{theoexample}
	\begin{equation*}
	\begin{cases}
	~~ \displaystyle -\kappa \mu q = \partial_t h^a - \phi^a q^2 + \frac{\left( \partial_q h^a \right)^2}{4k} & \qquad \qquad \text{\color{red} \small  (HJB equation - optimality)}
	\\
	~~ \displaystyle H^a(T,x,S,q;\mu) = x + q (S - \alpha^a q) & \qquad \qquad \text{\small \color{red} (HJB terminal condition)}
	\\
	~
	\\
	~~ \displaystyle \partial_t m + \partial_q  \bigg( m \cdot \frac{\partial h^a(t,q)}{2k} \bigg) = 0 & \qquad \qquad \text{\color{red} \small  (FP equation - density flow)}
	\\
	~~ \displaystyle m(0,q,a) = m_0(q,a) & \qquad \qquad \text{\color{red} \small  (FP initial condition)}
	\\
	~
	\\
	~~ \displaystyle \mu_t = \int_{(q,a)} \frac{\partial h^a(t,q)}{2k} ~m(t,dq,da) & \qquad \qquad \text{\color{red} \small  (net trading flow)}
	\end{cases}
	\end{equation*}
	
	\vspace{0.5cm}	
	\underline{Solution:} ~~~ see \cite{cardaliaguet2017mean}. 
\end{mytheo} 
\vspace{0.5cm}	

The main challenge with the MFG problem is that it involves both an HJB equation and a Fokker-Planck equation. Furthermore, the density governed by Fokker-Planck equation must remain positive on its domain and integrate to unity as we previously saw. The na\"{i}ve implementation of the MFG problem yields poor results given that the integral term $\mu_t $ is expensive to compute and the density in the Fokker-Planck equation has some constraints that must be satisfied. Using the same idea of exponentiating and normalizing used in \hyperref[sec:FK]{Section  \ref{sec:FK}}, we rewrite the density $m(t,q,a) = \frac{1}{c(t)} e^{-u(t,q,a)}$ to obtain a PDE for the function $u$:
\begin{equation*}
-\partial_t u + \frac{1}{2 k} (- \partial_q u \partial_q v + \partial_{qq} v) + \frac{\int (\partial_t u) e^{-u} dx}{\int e^{-u} dx} = 0
\end{equation*}
Both integral terms are handled by \textit{importance sampling} as in the Fokker-Planck equation with exponential transformation. \\

The equation was solved numerically with parameters $A,\phi,\alpha,k = 1$, with terminal time $T = 1$. 
The initial mass distribution was a normal distribution with mean $E_0 = 5$ and variance $0.25$. 
Results were calculated for $t \in [0,1]$ and $q \in [0,10]$.
The value function, optimal control along with the expected values of the mass through time were compared with the analytical solution (an analytical solution for the probability mass is not available; however the expected value of this distribution can be computed analytically). \\

The analytical solutions for the value function and the optimal control were in an acceptable range for the problem, though it should be noted that for $t = 0$, the approximation diverges as $q$ grows, but still fits reasonably well. The implied expected value from the fitted density showed a very good fit with the analytical solution. The probability mass could not be compared with an analytical solution, but it's reasonable to believe that it is close to the true solution given the remaining  results. \\

\begin{figure}[H]
	\centering
	\includegraphics[width=0.9\textwidth]{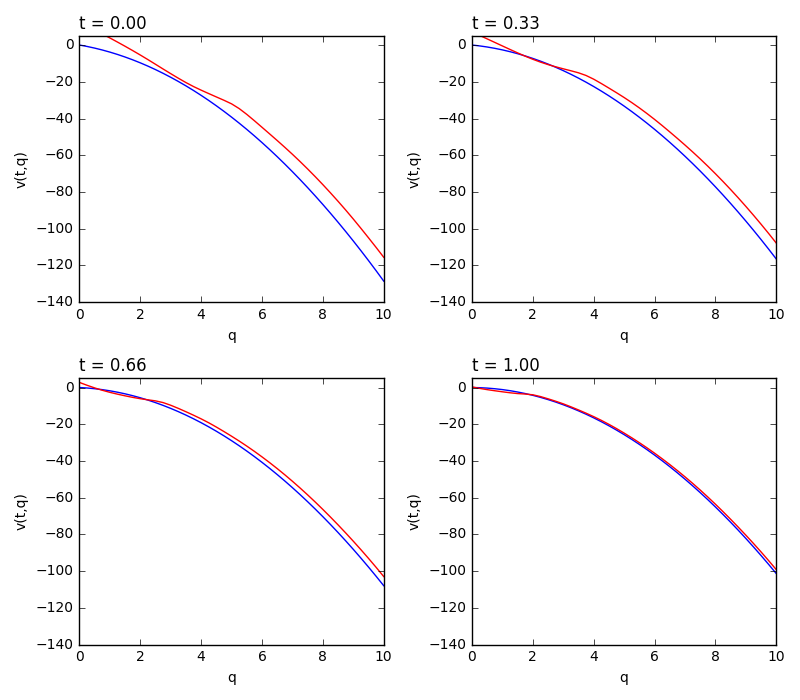}
	\caption{Approximate (red) vs. analytical solution for the value function of the MFG problem.}
\end{figure}
\begin{figure}[H]
	\centering
	\includegraphics[width=0.8\textwidth]{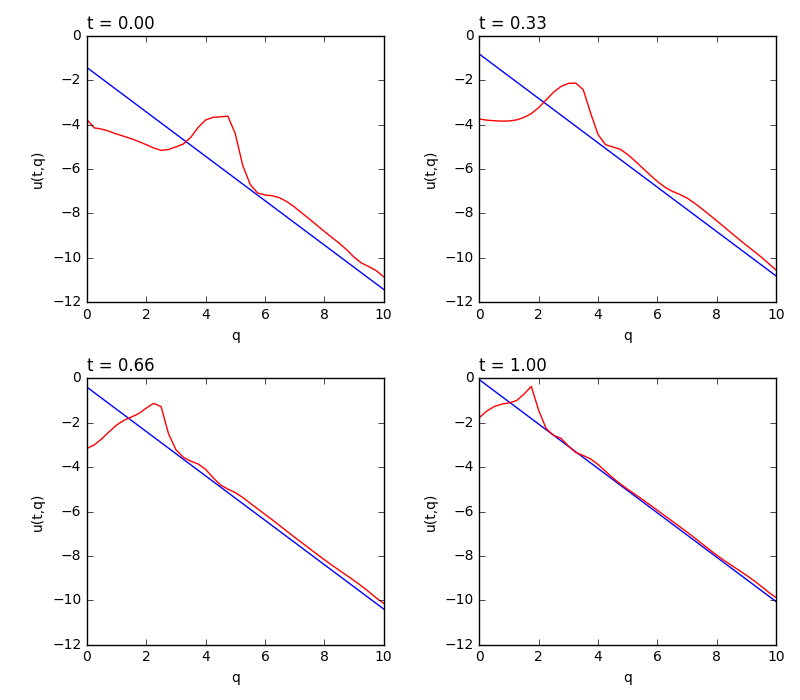}
	\caption{Approximate (red) vs. analytical solution for the optimal control for the MFG problem.}
\end{figure}

\begin{figure}[H]
	\centering
	\includegraphics[width=0.7\textwidth]{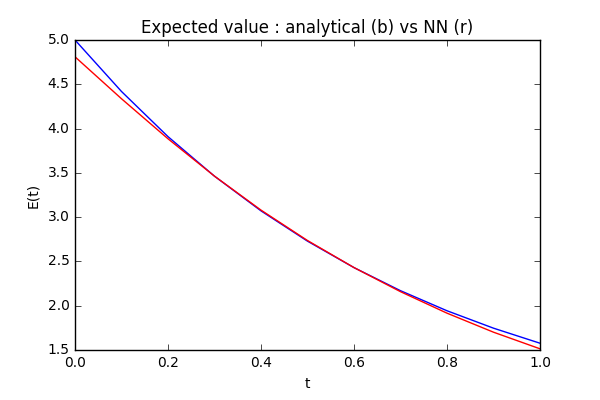}
	\caption{Approximate (red) vs. analytical solution for the expected value of the distribution of agents for the MFG problem.}
\end{figure}
\begin{figure}[H]
	\centering
	\includegraphics[width=\textwidth]{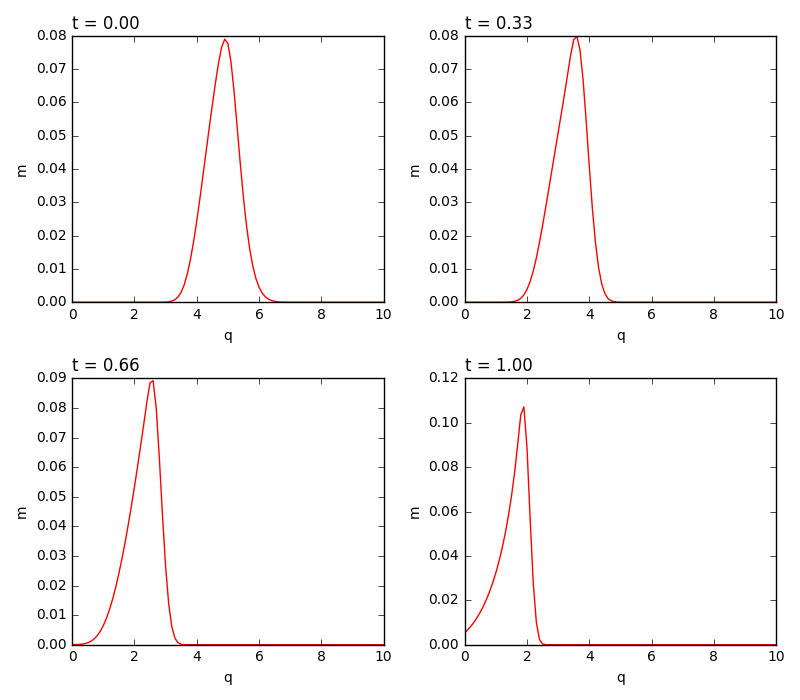}
	\caption{Unnormalized probability mass of inventories for MFG; the curve shifts left as all traders are liquidating.}
\end{figure}

\clearpage

\subsection{Conclusions and Future Work}

The main messages from the implementation of DGM can be distilled into \textbf{three main points}:
\begin{enumerate}
	\item \textbf{Sampling method matters}: similar to choosing a grid in finite difference methods, where and how the sampled random points used for training are chosen is the single most important factor in determining the quality of the results. 
	\item \textbf{Prior knowledge matters}: having some information about the solution can dramatically improve the accuracy of the approximations. This proved to be instrumental in the Fokker-Planck and MFG applications. It also rings true for finite difference methods and even Monte Carlo methods (a good analogy is the use of control variates).
	\item \textbf{Training time matters}: in some cases, including some of our earlier attempts, the loss functions appeared to be decreasing with iterations and the shape of solutions seemed to be moving in the right direction. As is the case with neural networks, and SGD-based optimization in general, sometimes the answer is to let the optimizer run longer. As a point of reference, \cite{sirignano2017dgm} ran the algorithm on a supercomputer with a GPU cluster and achieve excellent results in up to 200 dimensions. \\ 
\end{enumerate}

The last point regarding runtime is particularly interesting. While finite difference methods are memory-intensive, training the DGM network can take a long amount of time. This hints at a notion known in computer science as \textbf{space-time tradeoff}. However it should be noted that the finite difference approach will simply not run for high dimensionality, whereas the DGM (when properly executed) will arrive at a solution, though the runtime may be long. It would be interesting to study the space-time tradeoff for numerical methods used to solve PDEs. \\

As discussed earlier in this work, generalization in our context refers to how well the function satisfies the conditions of the PDE for points or regions in the function's domain that were not sampled in the training phase. Our experiments show that the DGM method does not generalize well in this sense; the function fits well on regions that are well-sampled (in a sense, this can be viewed as finding a solution to a similar yet distinct PDE defined over the region where sampling occurs). This is especially problematic for PDEs defined on unbounded domains, since it is impossible to sample everywhere for these problems using uniform distributions. Even when sampling from distributions with unbounded support, we may undersample relevant portions of the domain (or oversample regions that are not as relevant). Choosing the best distribution to sample from may be part of the problem, i.e. it may not be clear which is the appropriate distribution to use in the general case when applying DGM. As such, it would be interesting to explore efficient ways of tackling the issue of choosing the sampling distribution. On a related note, one could also explore more efficient methods of random sampling particularly in higher dimensions, e.g. quasi-Monte Carlo methods, Latin hypercube sampling. \\

Also, it would be interesting to understand what class of problems DGM can (or cannot) generalize to; a concept we refer to as \textbf{meta-generalization}. Is there an architecture or training method that yields better results for a wider range of PDEs? \\

Another potential research question draws inspiration from transfer learning, where knowledge gained from solving one problem can be applied to solving a different but related problem. In the context of PDEs and DGM, does knowing the solution to a simpler related PDE and using this solution as training data improve the performance of the DGM method for a more complex PDE? \\

Finally, we remark that above all neural networks are rarely a ``one-size-fits-all'' tool. Just as is the case with numerical methods, they need to be modified based on the problem. Continual experimentation and reflection is key to improving results, but a solid understanding of the underlying processes is vital to avoiding ``black-box'' opacity. \\

\clearpage

\subsubsection*{A Note On Performance}

In order to have a sense on how sensitive DGM is to the
computational environment used to train the neural networks, we benchmarked
training times both using and not using graphical processing units
(GPUs). It is well established among machine learning practitioners
that GPUs are able to achieve much higher throughput on typical neural
net training workloads due to parallelization opportunities at the
numerical processing level. On the other hand, complex neural network
architectures such as those of LSTM models may be harder to parallelize.
Some disclaimers are relevant at this point: these tests are not meant
to be exhaustive nor detailed. The goal is only to evaluate how much
faster using GPUs can be for the model at hand. Other caveats are
that we are using relatively small scale training sessions and we
are running on a relatively low performance GPU (GeForce 830M). \\

\subsubsection*{First test scenario}

Here we start with a DGM network with 3 hidden layers and 50 neurons
per layer. At first, we train the network as usual and then with no
resampling in the training loop to verify that the resampling procedure
is not significantly impacting the training times. The numerical values
are given in seconds per optimization step.

\begin{table}[H]
	\begin{centering}
		\begin{tabular}{|c|c|c|}
			\hline 
			Seconds / optimization steps & CPU & GPU\tabularnewline
			\hline 
			\hline 
			Regular training & 0.100 & 0.119\tabularnewline
			\hline 
			Training without resampling & 0.099 & 0.112\tabularnewline
			\hline 
		\end{tabular}
		\par\end{centering}
	\caption{In loop resampling impact}
	
\end{table}
Surprisingly, however, we also verify that the GPU is actually running
slower than the CPU! 

Next, we significantly increase the size of the network to check the
impact on the training times. We train networks with 10 and 20 layers,
with 200 neurons in both cases.

\begin{table}[H]
	\begin{centering}
		\begin{tabular}{|c|c|c|}
			\hline 
			Seconds / optimization steps & CPU & GPU\tabularnewline
			\hline 
			\hline 
			10 layers & 5.927 & 3.873\tabularnewline
			\hline 
			20 layers & 13.458 & 8.943\tabularnewline
			\hline 
		\end{tabular}
		\par\end{centering}
	\caption{Network size impact}
\end{table}

Now we see that the (regular) training times increase dramatically
and that the GPU is running faster then the CPU as expected. We hypothesize
that, given the complexity of DGM network architecture, the GPU engine
implementation is not able to achieve enough parallelization in the
computation graph to run faster than the CPU engine implementation
when the network is small.

\subsubsection*{Second test scenario}

We begin this section by noting that each hidden layer in the DGM
network is roughly eight times bigger than a multilayer perceptron
network, since each DGM network layer has 8 weight matrices and 4
bias vectors while the MLP network only has one weight matrix and
one bias vector per layer. So here we train a MLP network with 24
hidden layers and 50 neurons per layer (which should be roughly equivalent
with respect to the number of parameters to the 3 layered DGM network
above). We also train a bigger MLP network with 80 layers and 200
neurons per layer (which should be roughly equivalent with respect
to the number of parameters to the 10 layered DGM network above).

\begin{table}[H]
	\begin{centering}
		\begin{tabular}{|c|c|c|}
			\hline 
			Seconds / optimization steps & CPU & GPU\tabularnewline
			\hline 
			\hline 
			24 layers & 0.129 & 0.077\tabularnewline
			\hline 
			80 layers & 5.617 & 2.518\tabularnewline
			\hline 
		\end{tabular}
		\par\end{centering}
	\caption{Network size impact}
\end{table}

From the results above we verify that the GPU has a clear performance
advantage over the CPU even for small MLP networks.

We also note that, while the CPU training times for the different
network architectures (with comparable number of parameters) were
roughly equivalent, the GPU engine implementation is much more sensitive
to the complexity of the network architecture.